\documentclass[final,numbers,3p,12pt,times]{elsarticle}
\usepackage{amssymb,bm,amsmath}
\usepackage[nodots]{numcompress}

\biboptions{sort&compress}
\biboptions{square,numbers}

\journal{Computers and Fluids}

\begin{document}

\begin{frontmatter}

\title{A wavelet based numerical simulation technique for the two-phase flow using the phase field method}
\author[ref1]{Jahrul M Alam\corref{cor1}}
\ead{alamj@mun.ca}
\cortext[cor1]{Corresponding author}

\address[ref1]{Department of Mathematics and Statistics, Memorial University, Canada, A1C 5S7}

\begin{abstract}
  In multiphase flow phenomena, bubbles and droplets are advected, deformed, break up into smaller ones, and coalesce with each other. A primary challenge of classical computational fluid dynamics (CFD) methods for such flows is to effectively describe a transition zone between phases across which physical properties vary steeply but continuously. Based on the van der Walls’ theory, Allen-Cahn phase field method describes the face-to-face existence of two fluids with a free-energy functional of mass density or molar concentration, without imposing topological constraints on interface as phase boundary. In this article, a CFD simulation methodology is described by solving the Allen-Cahn-Navier-Stokes equations using a wavelet collocation method. The second order temporal accuracy is verified by simulating a moving sharp interface. The average terminal velocity of a rising gas bubble in a liquid that is computed by the present method has agreed with that computed by a laboratory experiment. The calculation of the surface tension force by the present method also shows an excellent agreement with what was obtained by an experiment. The up-welling and down-welling disturbances in a Rayleigh-Taylor instability are computed and compared with that from a reference numerical simulation. These results show that the wavelet based phase-field method is an efficient CFD simulation technique for gas-liquid or liquid-liquid flows.
%% Text of abstract
\end{abstract}


\end{frontmatter}

\section{Introduction}
The study of multiphase flow has attracted the attention of numerous researchers for well over a century because of its complexity and importance in areas of science and engineering~\cite{Fox2012,Rahman2013,Jeffrey2012,Wang2015}, such as Earth science, reservoir engineering, and geophysical fluid dynamics involving sediment, cloud, air pollution etc~\cite{Drew83}. There is thus a growing interest on the buoyant bubbles in a liquid, which plays a prototype model of multiphase flow systems~\cite{Bala2010,Fox2012}. For example, plugging up pipeline/wellbore usually causes a huge loss for petroleum industries due to production shutdowns~\cite{Rahman2015}. To develop an effective commercial design for the transportation of multiphase flows in such pipelines/wellbores one requires a better understanding of the surface tension, frictional pressure loss, and the operating condition necessary to avoid particle deposition and hydrate formation in pipelines and wellbores~\cite{Rahman2013}. 
In principle, pressure loss in a wellbore flow may be characterized by studying the dispersed multiphase flow regime. \citet{Bala2010} provides a detailed review of dispersed multiphase flow. In such a flow accurate representation of the fluid-fluid interface in a bubbly flow has motivated for the development of few popular multiphase flow modelling techniques~\cite{Fox2012,Joshi2015,Adler88,Sudheer2014}. 
The most utilized is the sharp interface approach which includes the volume of fluid~(VOF) method~\cite{Hirt81}, the level set~(LS) method~\cite{Osher88,Sethian2003}, the front tracking method~\cite{Tryggvason2001}, and the boundary integral method~\cite{Hou2001}. Such methods assume that the interface between two fluids has a zero thickness. The topological change of such an interface is tracked numerically based on the hyperbolic conservation law, and fluid properties on the interface are taken to be discontinuous~\cite{Tryggvason2001,Coutinho2014}.

In recent years, the extension of the phase-field method for modelling multiphase flow problems has also attained remarkable popularity. In contrast to the classical sharp interface methods, the popularity of the phase-field method is primarily due to its solid theoretical foundation and powerful capability of handling the topological change of fluid-fluid (or solid-fluid) interfaces. The PF method have been successfully applied to simulate bubbles rising in a liquid~\cite{Huang2014,Wang2015}, Rayleigh-Taylor instability~\cite{Ding2007}, thermo-capillary flows~\cite{Jasnow96}, droplet interaction~\cite{Zhang2010}.
In the PF methods, however, conserving mass for each phase is a critical issue. Although the continuum equations of the PF method can be modified to conserve mass, due to numerical dissipation introduced in discretization of the convective term, the total mass of the binary fluid is usually not conserved by the upwind method~\cite{Feng2007,Wang2015}. Such a numerical challenge was discussed in details by some researchers~\cite{Huang2014,Zheng2014,VDSman2008}. In the simulation of a bubble rising in a liquid, \citet{Huang2010} observed that the volume of a bubble can be decreased by $75$\% in an extreme case. To the best of knowledge, the mass non-conserving issue is primarily associated with the numerical scheme because it is convenient to apply a high-order upwind scheme to get a stable solution. The numerical dissipation of upwind schemes gets even worse for multiphase flows with a large density ratio.

In this work, we present a wavelet based mass-conserved PF method for effective simulation of multiphase flow problems. Our continuum model consists of the mass conserving Allen-Cahn phase field equations and the Navier-Stokes equation~\cite{Coutinho2014}. Based on the calculation of the residual of the Allen-Cahn equation, \citet{Coutinho2014} demonstrated that the mass of each phase can be conserved sufficiently accurately, using a Lagrange multiplier to modify the Allen-Cahn equation~\cite{Yang2006,Zhang2007}. The present author has extended the weighted residual collocation methodology of~\citet{Bruce72} for solving the Allen-Cahn phase field equation, where the residual has been projected onto a multiscale wavelet system at each time step~\cite{Mallat89,Mallat2009}. Our aim is to develop a wavelet based computational model for the fluid-fluid interface using the phase field approach. In wavelet methods, instead of using the upwind biased higher order explicit discretization, the solution is obtained iteratively using a wavelet projection algorithm. As described by~\citet{Alam2014,Alam2015b}, the discretization error of the present wavelet method is $\mathcal O(\Delta x^{2p})$, where $2p$ denotes vanishing moment of Deslauriers-Dubuc fundamental function used for the present wavelet system. Along with a second order discretization in time and using the heuristic stability analysis method presented by~\citet{Warming74}, it is easy to show that numerical dissipation with respect to the upwind method is absent in the present wavelet method. 

In the present computational approach, we take advantage of $i)$ a wavelet based weighted residual collocation method for the best approximation of multiphase flows, $ii)$ the phase field method to model the surface tension and the fluid-fluid interaction, and $iii)$ a Newton-Krylov-Wavelet method to computationally address multiphysics phenomena inherent into multiphase flow problems. In the literature, many advanced features of the multi-resolution wavelet method were investigated by several authors~\cite{Kai2010,Alam2012,Li2015}. Here, we investigate a simple but efficient wavelet method for simulating a single bubble rising in a liquid. We think that this is an important step in the direction of extending wavelet methods for multiphase flow simulations. Phase field method is a promising continuum description of fluid-fluid interactions. We think that discretizing the phase field equation based on multi-resolution wavelets is a natural extension for better computation of fluid-fluid interfaces. Over a decade the Newton-Krylov method is being advanced for effective simulation of multiphysics problems, but the approach is not fully understood for solving multiphase flow dynamics. Since the present methodology has attempted to take advantage of three important modern computational techniques, it is difficult to provide a complete details of each of the techniques. Our aim is to provide only a brief outline of the methodology and investigate its performance based on experimental and numerical data that are available from literature. Rest of the article thus focuses on presenting the method and discussing the primary results.

\subsection{Plan}

In section~\ref{sec:meth}, the phase field method and the wavelet based discretization of the governing phase field differential equations~(PDEs) are presented briefly. In section~\ref{sec:rva}, experimental validation of the proposed method is discussed. With respect to the CFD simulation, some experimental data are presented in this section. It is important to note that the accuracy of the surface tension calculation based on the phase field method is compared with experimental data. The numerical simulation of the Rayleigh-Taylor instability is compared with a reference numerical simulation. The article concludes with some remarks in section~\ref{sec:rmrk}.

\section{Methodology}\label{sec:meth}
\subsection{Allen-Chan phase-field method}
The phase field method is based on a free energy functional that describes the transition from on phase to another~\cite{Walls79}. A portion of the free energy is derived from the hydrophilic interaction, and the other is derived from the hydrophobic interaction. The competition between these two energy is resolved by a continuous phase variable $c(\bm x,t)$ the dynamics of which is generally described by either the Allen-Cahn~\cite{Allen79} or the Cahn-Hilliard~\cite{Cahn58} diffusion equation. 

In the present Allen-Cahn phase field method,  $c=0$ represents one phase ({\em e.g.} liquid), $c=1$ represents the other phase ({\em e.g.} gas), and the interface between two phases is represented by $0< c< 1$. In the phase field theory, the interface has a small but finite thickness~($\epsilon$), inside which two fluids may  mix and store a mixing energy that has a hydrophilic and hydrophobic contribution~\cite{Jacqmin99,Coutinho2014}. The `hydrophilic' effect tends to a complete mixing within the interface, representing a non-local interactions between two phases, which can be viewed as a classical Fick's law of diffuse interface. The `hydrophobic' effect tends to minimize the interfacial bulk free energy, which can be viewed as a classical sharp interface limit~\cite{Jacqmin99,Karma2002,Coutinho2014}. 

%The variational derivative of the total free energy is called the `chemical potential'~($\mu^c$). When the time scale~($1/\mathcal M$) for the elastic mixing of interfacial free energy tends to infinity, the phase field model approaches to the classical sharp interface model. In the limit of zero chemical potential ($\mu^c=0$), the interfacial mixing is in the equilibrium, and the capillary width is a measure of the diffuse interface.

Following~\citet{Ding2007}, let us consider a binary fluid that consists of two immiscible fluids, and define volume averaged quantities such as the density, the viscosity, and the viscous stress  of the binary fluid by
$$
\rho = \rho_gc+(1-c)\rho_l,\quad
\mu = \mu_gc+(1-c)\mu_l,\hbox{ and }
\tau_{ij} = \mu\left(\frac{\partial u_i}{\partial x_j} + \frac{\partial u_j}{\partial x_i}\right),% -\frac{1}{3}\delta_{ij}\frac{\partial u_k}{\partial x_k},
$$
respectively, where subscript $g\, (l)$ represents gas (liquid), and $u_i$ denotes the volume averaged velocity field. We then solve the Navier-Stokes equation and the Allen-Cahn phase field equation. %It is worth mentioning that viscosity for a three-phase flow is also calculated as $\mu = \mu_l(1+2.5c+10.06c^2+0.00273e^{16.6c})$, which is not used in the present article.

%The last term on the right hand side of $\tau_{ij}$ is absorbed into the modified pressure $p$, and the Boussinesq approximation is adopted when~(\ref{eq:nse}) is solved. 
\subsubsection{Navier-Stokes equations}
The results of~\citet{Antanovskii95} indicate that the conservation of momentum and mass for the phase field method may be expressed by the Navier-Stokes system~\cite{Ding2007,Coutinho2014,Wang2015}
\begin{equation}
  \label{eq:rho}
  %\frac{\partial p}{\partial t} + u_j\frac{\partial p}{\partial x_j} + p\frac{\partial u_j}{\partial x_j}=0.
  \frac{\partial\rho}{\partial t} +  u_j\frac{\partial\rho}{\partial x_j} + \rho\frac{\partial u_j}{\partial x_j}=0.
\end{equation}
\begin{equation}
  \label{eq:nse}
  \rho\left(\frac{\partial u_i}{\partial t} + u_j\frac{\partial u_i}{\partial x_j}\right) = -\frac{\partial p}{\partial x_i} + \frac{\partial\tau_{ij}}{\partial x_i} + f_i,
\end{equation}
In~(\ref{eq:nse}), both the body force $(g\rho\delta_{i2})$ and the surface tension force $(f_i^{\hbox{st}})$ are included in the last term such that $f_i=g\rho\delta_{i3}+f_i^{\hbox{st}}$. As discussed by~\citet{Antanovskii95,Ding2007}, if the diffusive mass flow of gas is equal and opposite to that of liquid,  eq~(\ref{eq:rho}) may be expressed as a divergence free condition $\frac{\partial u_j}{\partial x_j}=0$, which is also equivalent to adopting the Boussinesq approximation~\cite{Ding2007,Coutinho2014}.
%
%Eq~(\ref{eq:rho}) is derived to satisfy the conservation of mass so that the nonhydrostatic pressure~$(p)$ is related to the density~$(\rho)$ by $\frac{\partial p}{\partial\rho} = c^2_s$, where $c_s$ is the speed of sound~\cite{Alam2014,Tannehill97}. 
It is worth mentioning that a numerical difficulty in solving~(\ref{eq:nse}) lies in satisfying the conservation of mass~(\ref{eq:rho}). For an incompressible flow, the divergence free continuity equation may be regarded either as a constraint to determine the pressure, or the pressure plays the role of a Lagrange multiplier to satisfy the continuity equation. The pressure in~(\ref{eq:nse}) can be obtained by solving a Poisson equation, which is the most costly step~\cite{Alam2014,Tannehill97}. 

Alternatively, \citet{Chorin67} introduced the method of artificial compressiblity to surmount the difficulty of the incompressible limit. Over the years, a good success with the artificial compressibility method has been reported~\cite{Tannehill97}. As discussed by~\citet{Perrin2006}, one may start with~(\ref{eq:rho}), and employ the equation of state $p=c_s^2\rho$ so that~(\ref{eq:rho}) is replaced with~(\ref{eq:p})~\cite{Alam2014}
\begin{equation}
  \label{eq:p}
  \frac{\partial p}{\partial t} + u_j\frac{\partial p}{\partial x_j} + p\frac{\partial u_j}{\partial x_j}=0,
\end{equation}
where $c_s$ is the speed of sound.
A good success of this approach for simulating incompressible flows is discussed by~\citet{Perrin2006} and~\citet{Kundu2004}. Recently, investigators have verified the successful usage of~(\ref{eq:p}) for the simulation of two-phase flows using the Chan-Hilliard phase-field method~\cite{Lee2010,Wang2015,Wang2015b}. Here, we investigate on the Allen-Chan phase-field method using (\ref{eq:nse})~and~(\ref{eq:p}).

\subsubsection{Allen-Chan equation}
\label{sec:eva}
In the present work, a free energy density model for immiscible isothermal two phase fluids is considered, which is based on the phase field variable $c(\bm x,t)$ and it gradient. The coefficient of surface tension is then determined from an energetic variational principle using the modified Allen-Cahn phase field equation,
\begin{equation}
  \label{eq:c}
  \left(\frac{\partial c}{\partial t} + u_j\frac{\partial c}{\partial x_j}\right) = \mathcal M\mu^c,
\end{equation}
where $1/\mathcal M$ is a time scale for the elastic mixing of interfacial free energy, and $\mu^c$ is called the `chemical potential'~\cite{Jacqmin99,Coutinho2014,Wang2015}.
The thermodynamic free energy of a gas-liquid system consists of two portions -- a portion of which is responsible for the excess free energy in the interface between two phases, which is proportional to $\frac{1}{2}|\bm\nabla c|^2$, and the remaining portion is the bulk free energy that is usually treated through a double well potential or bulk entropy density ($f(c)$). The variational derivative of the total free energy
$$
\mathcal F(c) = \int_{\Omega} \left[f(c)+\frac{\epsilon^2}{2}|\bm\nabla c|\right]dV
$$
is known as the chemical potential
%$\mathcal F(c)$ is called chemical potential The potential in~(\ref{eq:c}) is given by $\mu^c=-\frac{\delta\mathcal F}{\delta c}$, where 
$$
\mu^c = -\frac{\delta\mathcal F}{\delta c} = \epsilon^2\nabla^2c - f'(c). 
$$
As it was done by~\citet{Ding2007}, we consider the bulk free energy density
$$
f(c) = c^2(1-c)^2, % + c^3(6c^2-15c+10).
%f(c) = \overbrace{c^2(1-c)^2}^{f_1(c)} + \eta_0\underbrace{c^3(6c^2-15c+10)}_{f_2(c)},
$$
which vanishes away from the interface. In order to ensure that $\frac{d}{dt}\int_{\Omega}c(\bm x,t)dV = 0$, a Lagrange multiplier may be introduced in the classical Allen-Chan model~(\ref{eq:c}), which leads to the additional term $\frac{1}{|\Omega|}\int_{\Omega}f(c)dV$ on the right hand side of~(\ref{eq:c})~\cite{Coutinho2014}. \citet{Wang2015} introduced similar modification of Chan-Hilliard equations in order to globally conserve mass. 
%
%
%where $\eta_0$ is a Lagrange multiplier in order to ensure that $\frac{d}{dt}\int_{\Omega}cd\bm x=0$~(see~\cite{Coutinho2014}). %The thickness of the interface is a balance between two opposing phenomena. The sharper the interface, the larger the amount of bulk free energy. The interface tends to diffuse in order to reduce the gradient free energy.
For brevity, the surface tension force can be stated using dimensionless quantities,
$$
f_i^{\hbox{st}} = \frac{\mu^c}{\mathcal Re\mathcal Ca}\frac{\partial c}{\partial x_i}.
$$ 
\citet{Ding2007} and \citet{Jacqmin99} provide further mathematical details for deriving the surface tension force. Here, the Reynolds number~($\mathcal Re$) and the Capillary number~($\mathcal Ca$) are defined, respectively, by
$$
\mathcal Re = \frac{\rho_lWD}{\mu_l},\quad\mathcal Ca=\frac{\mu_lW}{\sigma}.
$$
The characteristic scales for the velocity~($W$) and the length~($D$), and the surface tension force~($\sigma$) are parameters that are associated with an individual application.
Following sections provide an overview of the wavelet based numerical method.

\subsection{Computational methodology and setup}\label{sec:cm}
%\subsection{Wavelet based Newton-Krylov methods}\label{sec:cm}
% 
This article has investigated a wavelet based Newton-Krylov methodology for solving the phase field equations (AC-NSE system). The wavelet collocation method introduced by~\citet{Bertoluzza96,Oleg97} has been extended in the present work. In~\cite{Bertoluzza96,Oleg97}, derivatives are approximated by standard finite difference methods. An advantage of this approach is that one can use higher order upwind schemes along with some benefits of the multi-resolution wavelet method. However, the present wavelet collocation method is a special case of the weighted residual method~\cite{Bruce72}, where differential operators are approximated with a wavelet basis. This requires the implementation of the iterative sub-division scheme~\cite{Dubuc89} on two-(three-)dimensional meshes using the barycentric~\cite{Berrut2004} interpolation method~(e.g. \cite{Alam2014}). Based on the Deslauriers-Dubuc interpolating wavelets~\cite{Dubuc89,Alam2014}, the weighted residual of the nonlinear Allen-Cahn Navier-Stokes system is minimized with respect to the Krylov subspace. Let us express the system of multiphysics PDEs~(\ref{eq:nse}-\ref{eq:c}) with a compact notation $\mathcal L(u) =f $, where the fully coupled residual is $\mathcal R:= \mathcal L(u) -f $ is decomposed with following multi-resolution wavelet method.

%% In a compact notation, the discretized system of eqs~(\ref{eq:nse},\ref{eq:p},\ref{eq:c}) is written as
%% %
%% \begin{equation}
%%   \label{eq:bvp}
%%   \mathcal L(u) = f.
%% \end{equation}
%% %

\subsection{Method of weighted residual and Deslauriers-Dubuc (DD) Wavelets of order $2p-1$}
% 

%First, in the phase field method a critical issue is the numerical dissipation introduced by the explicit treatment of the advection terms in~(\ref{eq:c}). For example, numerical dissipation leads to the decrease of the mass of a drop or the volume of a rising bubble artificially. Sponteous shrinkage of a rising bubble reduced the bubble volume to zero due to accumulation of numerical error. 

%Second, the treatment of nonlinear terms implicitly requires matrix-vector multiplications at each time step, due to linearization of the nonlinear system, which has a computational cost $\mathcal O(\mathcal N^2)$, where $\mathcal N$ is the total number of the computational degrees of freedom.

%Based on the heuristic stability analysis, we have considered a wavelet based discretization so that numerical dissipation can be minimized without refining the mesh in contrast to classical schemes that require a fine mesh to reduce the numerical dissipation.
%
%Introduced in \cite{Bertoluzza96,Oleg97}, we have developed a wavelet collocation method to address the above challenges. This method solves a system of nonlinear PDEs, which can be written with the following compact notation (including the boundary conditions):

%We briefly present a simple but efficient wavelet based CFD method~\cite{Dubuc89,Kai2010,Alam2014}, where the phase-field equations are discretized using the principle of multiresolution algorithm developed by~\citet{Mallat89}. 

%\subsubsection{Deslauriers-Dubuc (DD) Wavelets of order $2p-1$}
%
\begin{figure}
  \centering
  \begin{tabular}{c}
    \includegraphics[height=6cm]{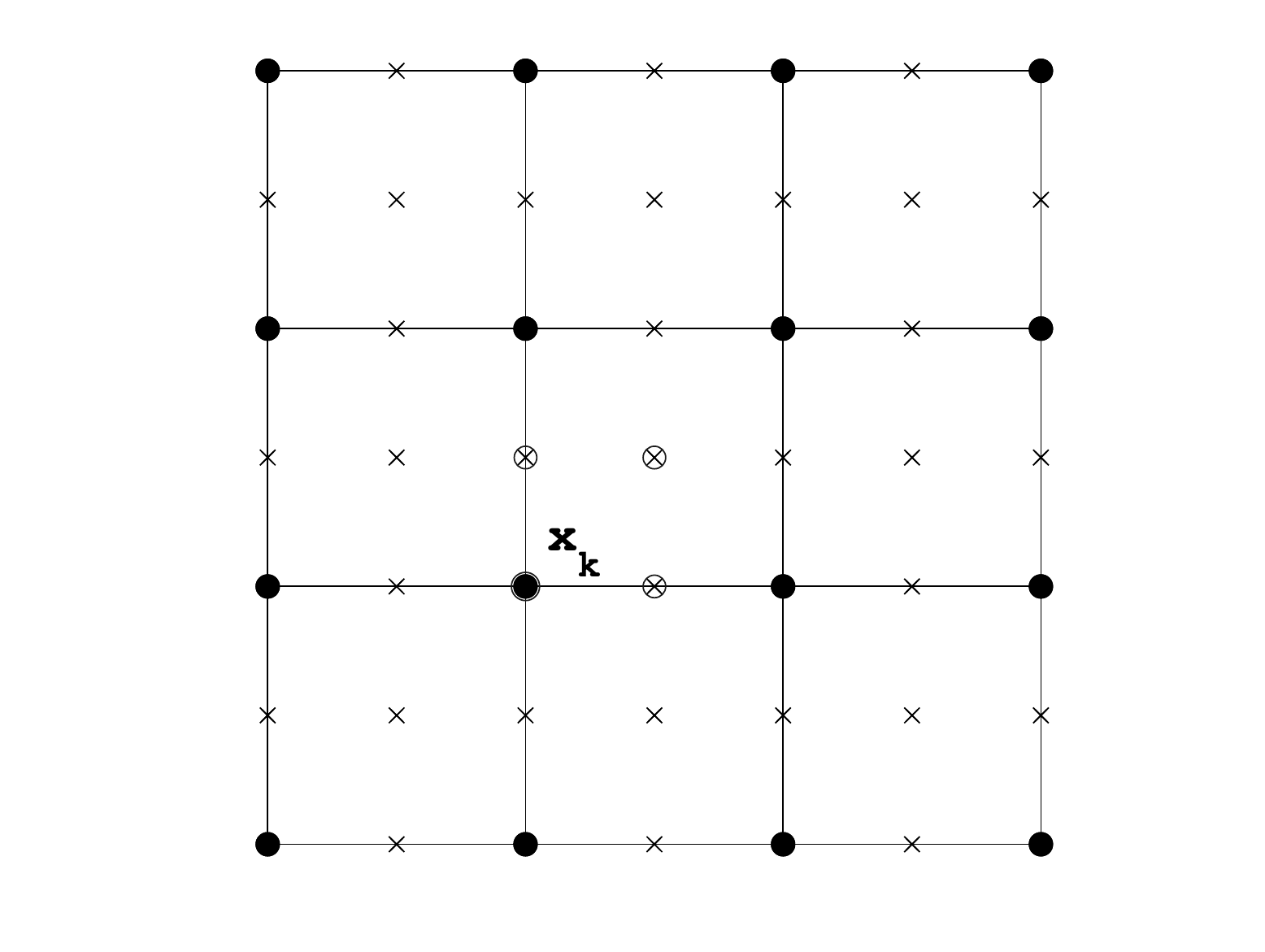}
  \end{tabular}
  \caption{A multi-resolution mesh $\mathcal G^1$, where all nodes in the coarse mesh $\mathcal G^0$ are denoted by `$\bullet$', and the additional nodes in the refined mesh $\mathcal G^1$ are denoted by `$\times$'. }
  \label{fig:msh}
\end{figure}

In order to model a fluid-fluid interface, the present wavelet method is constructed on a two-dimensional multi-resolution mesh that is a collection of rectangular computational cells. To illustrate the iterative subdivision scheme~\cite{Dubuc89}, a two resolution mesh $\mathcal G^1$ is shown in Fig~\ref{fig:msh}, where all nodes in the coarsest mesh $\mathcal G^0$ are denoted by~`$\bullet$'. In the refined mesh $\mathcal G^1$, all new nodes with respect to $\mathcal G^0$ are denoted by~`$\times$'. Given the values of a function on $\bullet$ nodes, the iterative sub-division scheme interpolates it on all $\times$ nodes. The process is repeated by successively refining the mesh to obtain a multi-resolution mesh $\mathcal G^s$ such that ({\em e.g.} Chapter 7.8 of~\cite{Mallat2009})
$$
\mathcal G^0\subseteq\cdots\subseteq\mathcal G^{s-1}\subseteq\mathcal G^s\subseteq\mathcal G^{s+1}\subseteq\cdots.
$$
%For example in Fig~\ref{fig:msh}, all nodes in the mesh $\mathcal G^0$ are denoted by~`$\bullet$'. In the refined mesh $\mathcal G^1$, all new nodes with respect to $\mathcal G^0$ are denoted by~`$\times$'. 
%Let $\varphi(\bm x)$ be a fundamental function develped by applying the DD method of~\cite{Dubuc89} on the mesh $\mathcal G^0$.
In the multi-resolution theory~\cite{Mallat89}, a basis $\{\varphi_k(\bm x)\}$ is defined with respect to the nodes $\bm x_k\in\mathcal G^0$. However, on the refined mesh $\mathcal G^1$, a new basis $\{\psi_k(\bm x)\}$ is defined with respect to new nodes  $\bm x_k\in\mathcal G^1\backslash\mathcal G^0$ (as marked by `$\times$' in Fig~\ref{fig:msh}) in order to capture additional details of a function with respect to the coarser level features. The process is repeated until a desired level of refinement is obtained. %\citet{Dubuc89} provides an efficient iterative algorithm for the multi-resolution decomposition based on $\varphi_k(\bm x)$ and $\psi_k(\bm x)$ (see also,~\cite{Donoho92,Alam2014}). 

Finally, if the refined mesh has $\mathcal N$ nodes, the residual $\mathcal R(u^{\mathcal N})$ is approximated with respect to a trial solution $u^{\mathcal N}(\bm x)$  representing a fluid flow, which has the following multi-resolution wavelet decomposition~\cite{Bertoluzza96,Alamphd2006,Kai2010})
\begin{eqnarray}
  \label{eq:mra}
  u^{\mathcal N}(\bm x) &=& \sum_{k=0}^{\mathcal N-1}c_k\varphi_k(\bm x),\\
  \label{eq:mrb}
  &=&\sum_{\bm x_k\in\mathcal G^0}^{}c_k\varphi_k(\bm x)
      +
      \sum_{s=1}^{\mathcal S}\sum_{\bm x_k\in\mathcal G^s\backslash\mathcal G^{s-1}}^{}d_k\psi_k(\bm x),
\end{eqnarray}
where the wavelet expansions (\ref{eq:mra}) and~(\ref{eq:mrb}) are equivalent. In principle, the residual $\mathcal R$ is decomposed by~(\ref{eq:mrb}) and the corresponding wavelet expansion coefficients are determined by the collocation method~\cite{Bruce72} so that $\mathcal R$ vanishes on all nodes $\bm x_k$ of the refined mesh. For brevity, we do not provide the technical details of determining the coefficients, which can be found from refs, {\em e.g.}~\cite{Alam2014,Alam2015b,Mallat2009}.
Clearly, scale-by-scale information contained in the solution $u(\bm x)$ representing a fluid-fluid interface is approximated by $u^{\mathcal N}(\bm x)$. Since the term $\sum d_k\psi_k$ is associated to additional nodes on a refined mesh with respect to a coarse mesh, it captures additional details between a coarse mesh and a refined mesh, which is an advantage of the multiresolution method presented by~\citet{Mallat89}. As a result, any interface between two fluids is approximated more accurately with the help of detailed wavelet basis associated with the additional nodes on a refined mesh. In the literature, several other aspects of the wavelet method is seen promising for simulating fluid flows having sharp interfaces~\cite{Kai2010,Alam2006,Alam2012,Alam2014}. To provide some interesting features of the wavelet method in the context of modelling the time evolution of a two-phase flow, it may be convenient to provide the following computational properties of $\varphi(x)$ using a one-dimensional setting.

\begin{enumerate}
\item $\varphi(x)$ vanishes outside the interval $[x_{-2p+1},x_{2p-1}]$ and has exactly $4p-2$ zeros within its support. As a result, the multi-resolution discretization provides a sparse representation of PDEs. Note that $p=3$ has been taken for most results of this article. 

\item The basis $\{\varphi_k\}$ reproduces polynomials up to a degree $2p-1$, which means that the conditioning of the discetization is influenced primarily by the interface.

\item $\varphi(x)$ is an autocorrelation of the Daubechies scaling function, where $p$ is the number of vanishing moments of the corresponding Daubachies wavelets. As discussed by~\citet{Bertoluzza96}, this property helps wavelet methods to get well-conditioned discretization.

\item The first derivative of $\varphi(x)$ vanishes at $x=0$ and outside the support of $\varphi(x)$, and is nonzero at every other zeros of $\varphi(x)$. \citet{Alam2014} extended this property to develop a wavelet based discretization technique which is $\mathcal O(\Delta x^p)$. 

\end{enumerate}

\subsubsection{The Newton-Krylov-Wavelet method}
Let us suppose that the fully coupled residual $\mathcal R$ is minimized by the method of weighted residual~\cite{Bruce72} based on the wavelet decomposition~(\ref{eq:mra})~\cite{AlamPhD,Alam2014,Knoll2004}. In the wavelet collocation method, the residual $\mathcal R$ vanishes on every nodes, which can be achieved iteratively so that $\lim\limits_{\mathcal N\rightarrow\infty}\mathcal R=0$ everywhere in the domain. In other words, the given system of PDEs is reduced to a system of coupled nonlinear equations
\begin{equation}
  \label{eq:lu}
  \mathcal R(u^{\mathcal N}) = 0.
  %\mathcal L(u^{\mathcal N}) = f.
\end{equation}
The system of eqs~(\ref{eq:lu}) is then solved to obtain the parameters $c_k$'s (or $d_k$'s), and the approximate solution is formed with~(\ref{eq:mra}) or~(\ref{eq:mrb}).
The idea of solving the simultaneously coupled nonlinear system is to capture the multiphysics phenomena accurately~\cite{Knoll2004}.
The solution of~(\ref{eq:lu}) is approximated iteratively such that $u^{\mathcal N}=\lim\limits_{k\rightarrow\infty} u^k$. The iteration $u^{k+1}=u^k+s^k$ is completed by the Newton method with respect to a multivariate Taylor expansion around a previous iterate $u^k$:
$$
\mathcal R(u^{k+1} + s^k)\approx \mathcal R(u^k) +   \mathcal J(u^k)s^k.
$$
Setting $\mathcal R(u^{k+1} + s^k) = 0$ the correction $s^k$ is obtained by solving the linear system of $\mathcal N$ equations:
\begin{equation}
  \label{eq:js}
  \mathcal J(u^k)s^k = -\mathcal R(u^k), %f-\mathcal L(u^k),
\end{equation}
where the Jacobian matrix ($\mathcal J$) is given by $\mathcal J_{ij} = \frac{\partial\mathcal R_i}{\partial u_j}$. 
In the literature, the coupled approach is often called the multiphysics solver in contrast to the segregated solver, where each physics is approximated individually. For the classical numerical schemes, there is a high computational cost of forming~(\ref{eq:js}) because of the matrix vector product requiring $\mathcal O(\mathcal N^3)$ operations at each $k$-th iteration. However, in wavelet collocation methods~\cite{Alam2012}, it is easy to accelerate the computation through the use of Frechet derivative of $\mathcal R(u)$~(see also~\cite{Knoll2004}), thereby bringing down the overall computational cost to $\mathcal O(\mathcal N)$. Let us estimate the Frechet derivative by
$$
\mathcal J(u^k)s^k\approx\frac{\mathcal R(u^k+s^k\eta)-\mathcal R(u^k)}{\eta},\quad\eta\rightarrow 0,
$$
which requires only $\mathcal O(\mathcal N)$ operations for the wavelet decomposition of $\mathcal R$~\cite{Knoll2004,Alam2012}. 
Although the present wavelet method always uses a symmetric stencil for discretization, due to the multiphysics nature of the multiphase flows, the system~(\ref{eq:js}) is often not symmetric and positive definite. In wavelet method, it is also easy to pre-condition the system~(\ref{eq:js}) so that a Krylov method is often an optimal solver. In particular, we have implemented the GMRES (generalized minimal residual) method for solving two-phase flow problems.
\subsubsection{Verification:  a single phase moving interface}\label{sec:burg2d}
To outline some of the features of the present method, let us consider the flow of an incompressible fluid in a two-dimensional box $[0,1]\times[0,1]$ without any solid boundaries. We can form an exact solution of~(\ref{eq:nse}) assuming that $\rho=1$, $\tau_{ij}=\mu\frac{\partial u_i}{\partial x_j}$, $f_i=0$, and $p=0$. Let the initial velocity field be given by,
$$
u_1(x,y,0) = \frac{3(1+\exp((-4x+4y)/(32\mu))-1)}{4(1+\exp((-4x+4y)/(32\mu)))}
$$
and
$$
u_2(x,y,0) = \frac{3(1+\exp((-4x+4y)/(32\mu))+1)}{4(1+\exp((-4x+4y)/(32\mu)))}.
$$
With these assumptions each component of the solution of~(\ref{eq:nse}) exhibits a moving interface. %We can use this manufactured solution to assess the numerical integration error. %Note that the treatment of the coupled system~(\ref{eq:nse}-\ref{eq:c}) is very similar to the following simplified description.
Let us now discuss the leading order truncation error employing a second order temporal integration scheme for eq~(\ref{eq:nse}) such that
$$
\underbrace{
  \frac{2 u_i^{n+1}}{\Delta t} + u_j^{n+1}\frac{\partial u_i^{n+1}}{\partial x_j}-\mu\frac{\partial^2 u_i^{n+1}}{\partial x_j\partial x_j} 
}_{\mathcal L(u^{\mathcal N})}
=
\underbrace{
  \frac{2 u_i^{n}}{\Delta t} - u_j^{n}\frac{\partial u_i^{n}}{\partial x_j} + \mu\frac{\partial^2 u_i^{n}}{\partial x_j\partial x_j} 
}_f.
$$
For the above scheme the leading order error contains only derivatives of order $2p+1$, and is $\mathcal O(\Delta t^2,\Delta x^{2p})$.
As the heuristic stability analysis method was discussed by~\citet{Warming74} and~\citet{Tannehill97}, the odd order derivatives in the leading order term of the error causes dispersive error, and the dissipative error is caused by the even order derivatives. Based on the heuristic method of~\citet{Warming74}, the above discretization is unconditionally stable, and does not produce dissipative errors. As a result, we expect that the moving interface of a finite thickness will not be smeared out or dissipated. 

%For simplicity, let us consider the local truncation error with respect to derivatives in the $x\,(=x_1)$ direction only and denote the corresponding $2p+1$-st order derivative by $u^{2p+1}$. The leading order error due to discretization of the advection terms is $\mathcal O(\Delta t^2,\Delta x^{2p},u^{(2p+1)})$ and that due to the diffusion terms is $\mathcal O(\Delta x^{2p},u^{(2p+2)})$~\cite{Tannehill97,Alam2014}. Note that the even and the odd derivative terms in the leading order error contribute for numerical dissipation and dispersion, respectively~\cite{Warming74,Tannehill97}. Clearly, the discretization of advection terms does not lead to numerical dissipation. Although the derivation is not rigorous, the above discussion demonstrates that the primary source of error is due to numerical dispersion. %  In contrast, to simulate such a moving interface, a classical approach would employ a dedicated advection algorithm such as an upwind or an essentially non-oscillatory~(ENO) finite difference/volume scheme~\cite{Harten97,Tannehill97}. However, the `artificial numerical dissipation' and the CFL~(Courant-Friedrichs-Lewy) stability restriction of these classical schemes make them inaccurate and costly. The accuracy of the proposed method is demonstrated with the following example.

\begin{figure}
  \centering
  \begin{tabular}{cc}
    \includegraphics[height=4cm]{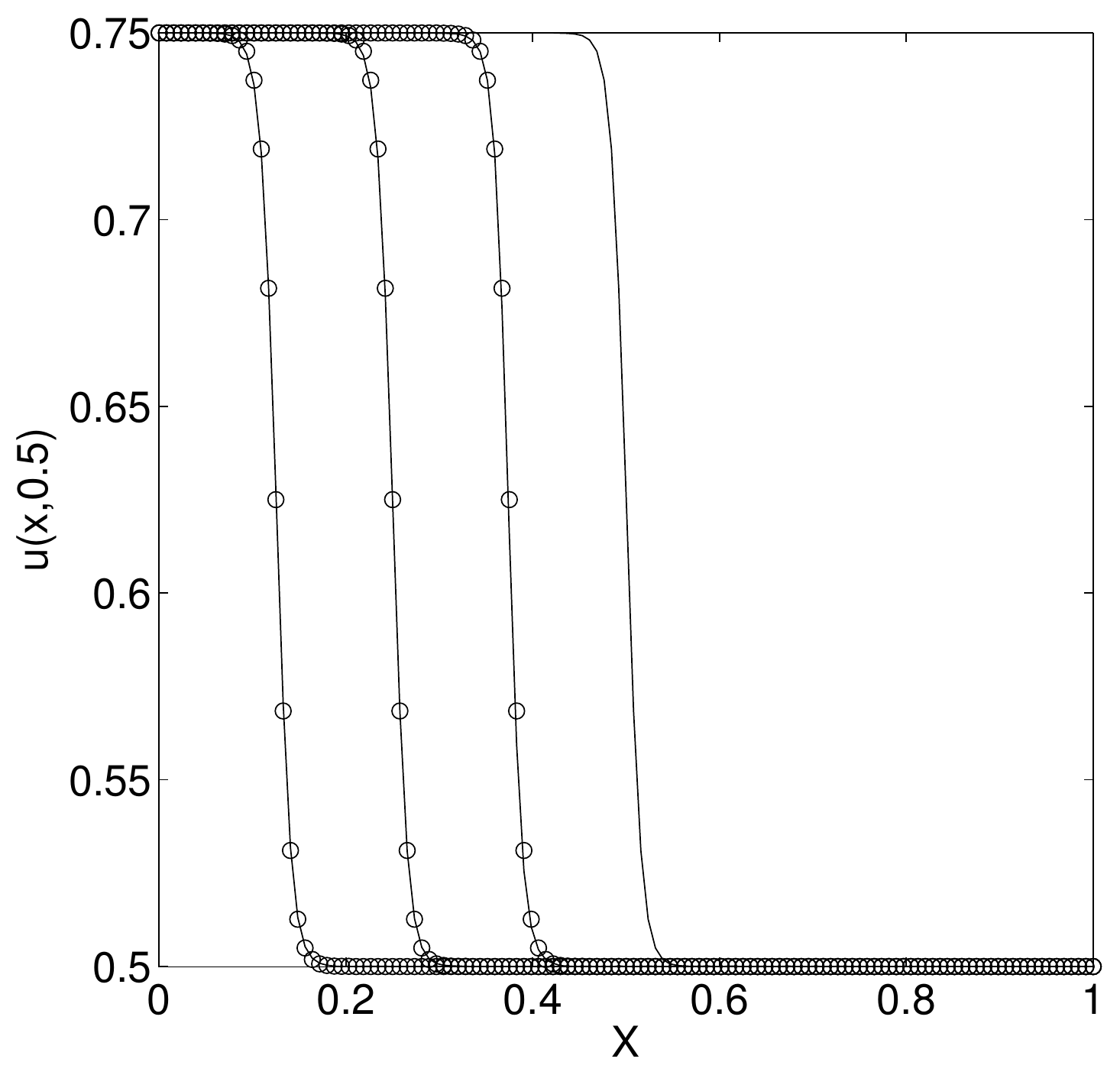}&
    \includegraphics[height=4cm]{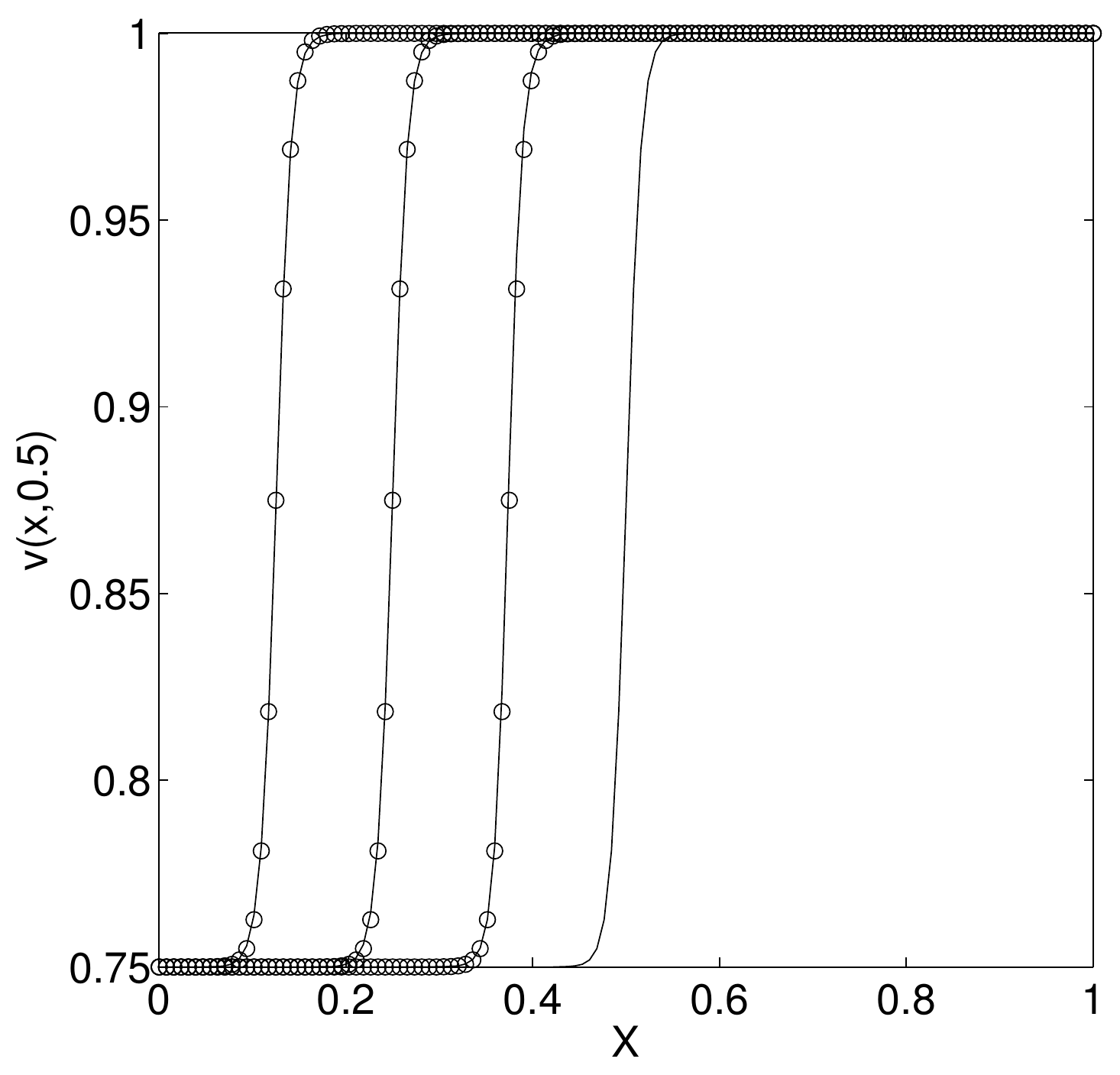}\\
    $(a)\, u_1(x,0.5)$ & $(b)\,u_2(x,0.5)$\\
    \includegraphics[height=4cm]{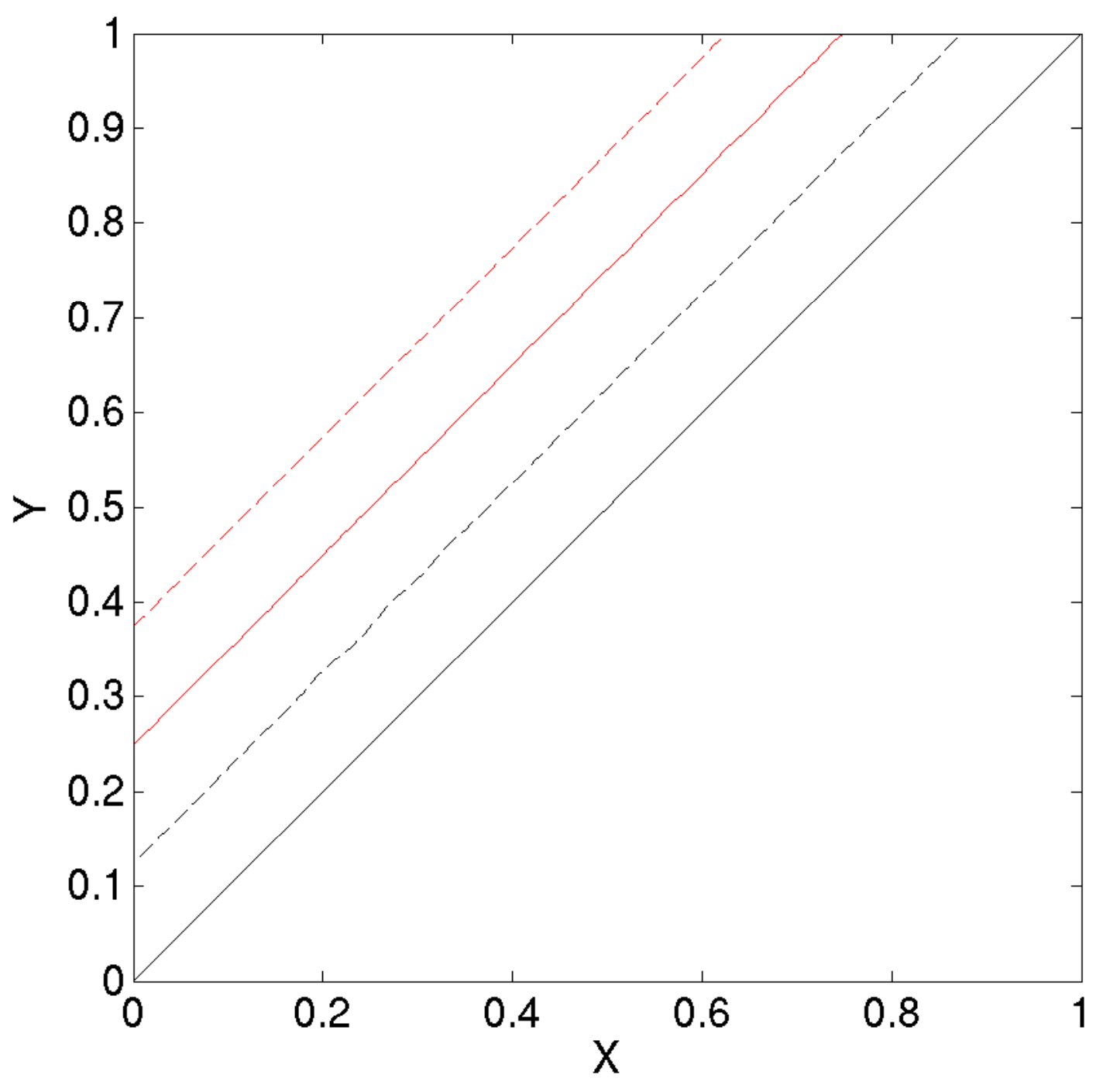}&
    \includegraphics[height=4cm]{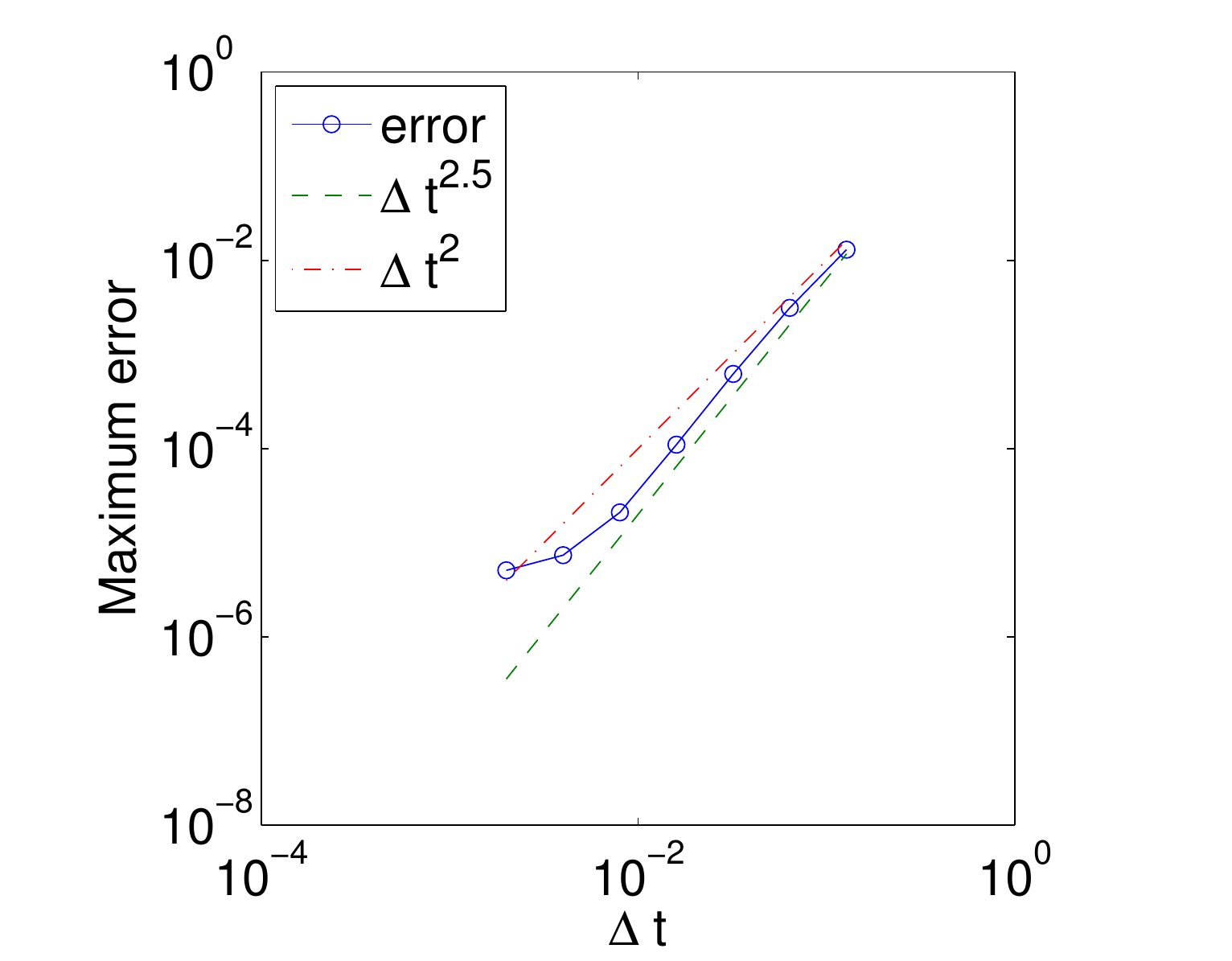}\\
    $(c)$ & $(d)$\\
    % \multicolumn{2}{c}{
    % \includegraphics[height=4cm]{plots/burg2derr}
    % }\\
    % \multicolumn{2}{c}{
    % $(c)$
    % }\\
  \end{tabular}
  \caption{$(a$-$b)$ The numerical approximations for the solutions of~(\ref{eq:nse}); solid line, numerical approximations to $u_1(x,0.5)$ and $u_2(x,0.5)$ at $t=0$, $0.5$, $1.0$, and $1.5$ ($t$ increases from right to left); symbol $(\circ)$, exact solution. $(c)$ The moving interface $u_1(x,y,t)=0.625$ at at $t=0$, $0.5$, $1.0$, and $1.5$. $(d)$ An estimate for the temporal integration error showing the overall second order accuracy.}
  \label{fig:brg}
\end{figure}

%\subsubsection{Numerical example}
%
Using a mesh with $\mathcal N=129\times 129$ nodes and a Courant-Friedrichs-Lewy number, CFL=$12.8$, moving interfaces of a finite thickness have been simulated, which are solutions $u_1(x,0.5,t)$ and $u_2(x,0.5,t)$ at dimensionless times $t=0,\,0.5,\,1.0,\,1.5$, and are shown in Fig~\ref{fig:brg}$(a,b)$. The numerical solutions (solid line) are compared with the exact solutions (symbol $\circ$). Clearly, an excellent accuracy is observed although a large value for the CFL number was used. In~\citet{Tannehill97}, it was demonstrated that most explicit finite difference methods exhibit oscillatory behaviour when a moving interface is simulated. Based on this information, one finds that the present wavelet method provides an accurate result without using any dedicated scheme to capture a moving interface.

The robustness of the present Newton-Krylov-Wavelet solver can be examined by estimating the rate of convergence with respect to time truncation error.
Since the method is unconditionally stable, the time step can be adjusted to meet a desired level of accuracy. Here, the maximum error accumulated at $t=1.5$ has been computed as a function of the time step $\Delta t$, and the result is shown in Fig~\ref{fig:brg}$d$. Clearly, these data confirms the second order accuracy of the method as expected.

\section{Results, validation, and application}\label{sec:rva}
\subsection{A gas bubble rising in a liquid}
%\subsection{Parameters}
The multiphase flow phenomena discussed in this article may be characterized by three dimensionless parameters. These parameters are $\mathcal Re=\frac{\rho_lWD}{\mu_l}$, $\mathcal Pe = \frac{WD}{\mathcal M\epsilon^2}$, and $\mathcal Ca=\frac{\mu_lW}{\sigma}$, where the characteristic length and the velocity scales are denoted by $D$ and $W$, respectively. Characteristic values of various physical parameters and the associated dimensionless parameters are given for each test case. \citet{Bhaga81} provides additional parameters -- E{\"o}tv{\"o}s number, Morton number, and Weber number -- in order to explain experimental results. \citet{Hua2007} presented bubble shapes as a function of the Reynolds number and the Capillary number. When $\mathcal Re=1$, a bubble retains its initial spherical shape for the Capillary numbers $0.5\le\mathcal Ca\le 200$. However, for $\mathcal Re > 1$, a rising bubble may deform quickly or break into multiple bubbles for $\mathcal Ca\ge 1$. The topological shape of bubbles is sensitive to $\mathcal Re$, $\mathcal Ca$, $\rho_g/\rho_l$, and $\mu_g/\mu_l$ (see fig 6 of~\cite{Hua2007}). \citet{Bhaga81} reported experimental observation of the terminal shape of rising bubbles as a function of the Reynolds number, Morton number, and E{\"o}tv{\"o}s number. Such dynamics of a gas bubble rising in a liquid has been studied for a century and continues to be a problem of great interest today in the field of multiphase flow~\cite{Tryggvason2001,Coutinho2014}.

%\subsection{A gas bubble rising in a liquid}
%
%The proposed method is tested by simulating the  dynamics of a gas bubble in a quiescent liquid because this is a benchmark test case that was studied by several researchers, both experimentally and numerically~\cite{Bhaga81,Hua2007}. Since the bubble dynamics vary greatly in various flow conditions,  understanding the bubbly flow regime is also an interesting topic to industries that are involved in oil and gas transportation. In general, small bubbles maintain a spherical shape when they rise in a liquid. However, the shape of larger bubbles are affected significantly by the flow conditions~\cite{Bhaga81}.  

\subsubsection{Interfacial dynamics of a gas bubble rising in a liquid}
\begin{table}[h]
  \centering
  \begin{tabular}{|l|l|}
    \hline
    Parameter & Value \\
    \hline
    $\rho_l$ [kg/m$^3$] & $10^3$\\
    $\rho_g$ [kg/m$^3$] & $1.225$\\
    $\mu_l$ [kg/ms] & $3.50\times 10^{-1}$\\
    $\mu_g$ [kg/ms] & $3.58\times10^{-3}$\\
    $\sigma$ [kg/s$^2$] & $1.10\times 10^{-1}$\\
    $g$ [m/s$^2$] & 9.81\\
    $D$ [m] & $5\times 10^{-2}$\\
    \hline
    \end{tabular}
  \caption{List of parameters for the result presented in Fig~\ref{fig:bbct}.}
  \label{tab:pp}
\end{table}
This section demonstrates a comparison between the present simulation and a reference numerical model presented by~\citet{Ding2007}.
Physical parameters for the transition of an axisymmetric spherical rising bubble has been listed in table~\ref{tab:pp}. The characteristic velocity scale is given by $W=\sqrt{gD\Delta\rho/\rho_l}$. The Reynolds ($\mathcal Re$) number and the Capillary $(\mathcal Ca$) number are $100$ and $2.228$, respectively. In order to help comparison, the values of $\mathcal Re$, $\mathcal Ca$, the viscosity ratio, $\mu_g/\mu_l$, and the density ratio $\rho_g/\rho_l$ are about the same as what were used by~\citet{Ding2007}.  A spherical bubble of radius $D/2$ is introduced in a cylindrical tank of radius $2D$ and height $4D$. Assuming spherical symmetry, the computational domain $2D\times 4D$ is a vertical cross section of a cylindrical tank, and the initial center of the bubble is placed at a height of $D$ above the bottom and along the axis of the tank. 
The simulation has been done in the time interval $[0,4T]$ with a fixed time step $\Delta t=0.025T$, where $T=D/W$ is the time scale. 

Using the Allen-Chan phase field method, the present wavelet based simulation has been compared with the Chan-Hilliard phase field simulation of~\citet{Ding2007}.
The convergence of the algorithm is verified with a set of results at $4$ different resolutions, $n_x\times n_z$: $33\times 65$, $65\times 129$, $129\times 257$, and  $257\times 513$, where $n_x$ and $n_z$ are the number of nodes in $x$ (radial) and $z$ (axial) directions, respectively. Based on the maximum vertical velocity, the Courant number (CFL) for these resolutions are $0.4$, $0.8$, $1.6$, and $3.2$, respectively. As discussed in the introduction, the multi-resolution wavelet method would improve the prediction of the interface between gas and liquid as the resolution $n_x\times n_y$ increases. 

\begin{figure}
  \centering
  \begin{tabular}{cc}
    \includegraphics[trim=0cm 0cm 0cm 0cm,clip=true,height=6cm]{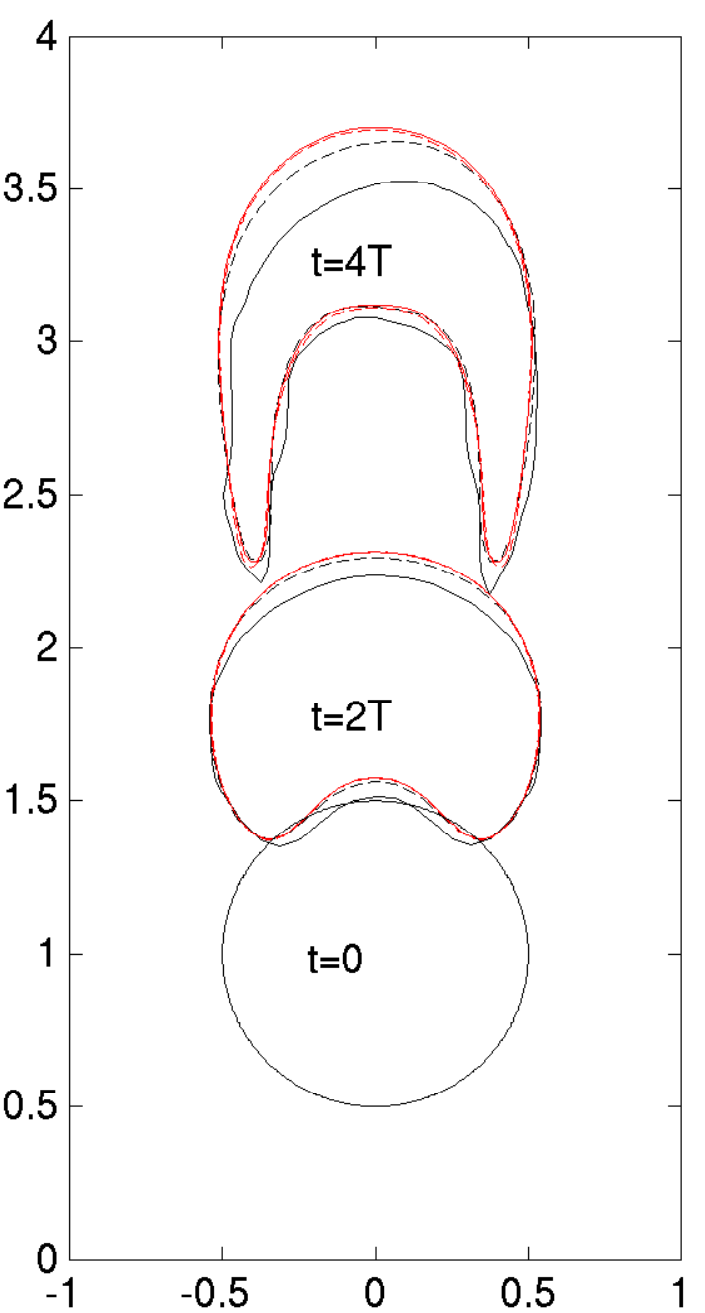}  
    &
    \includegraphics[trim=0cm 0cm 0cm 0cm,clip=true,height=6cm]{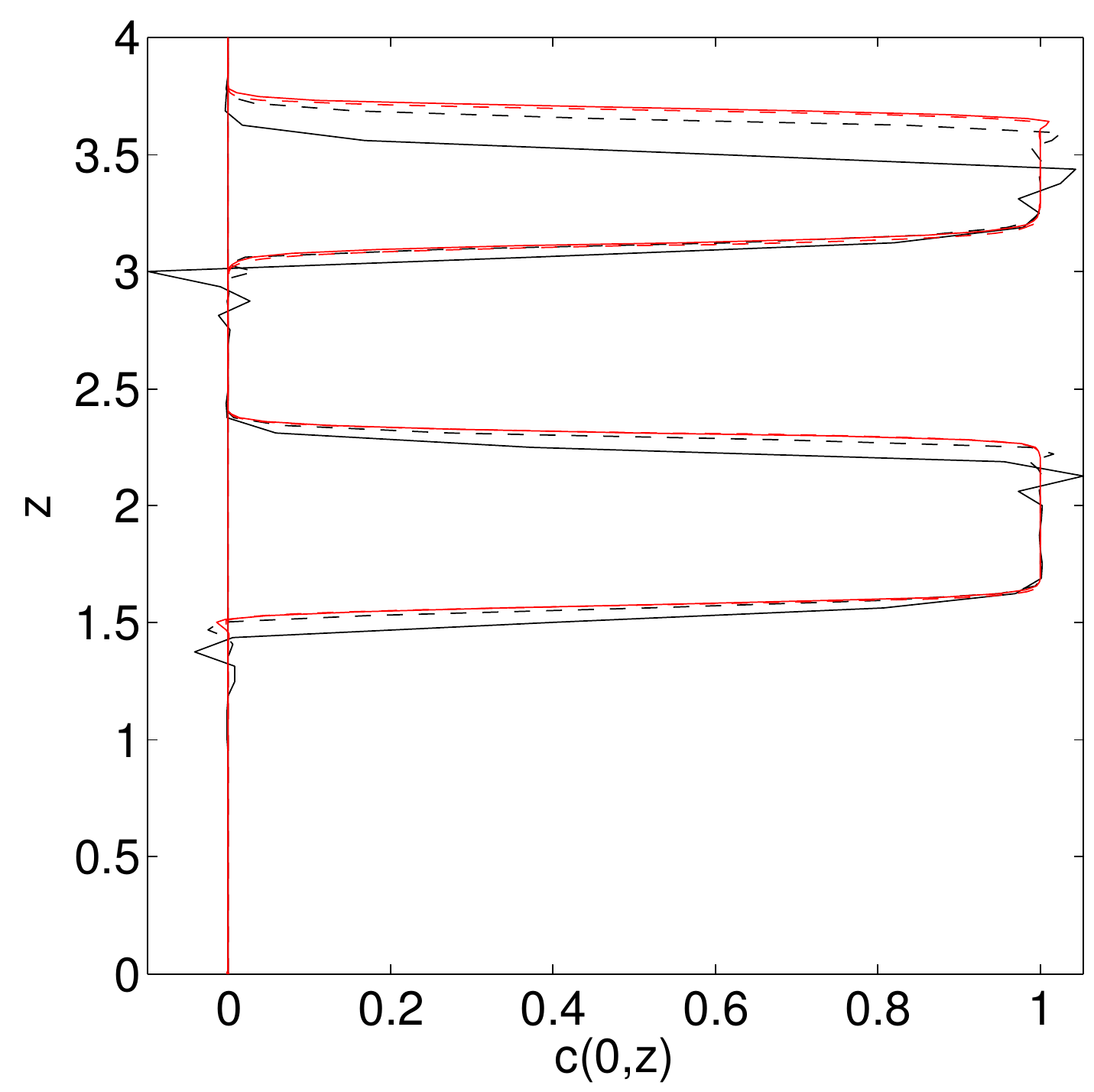}  
    \\
    $(a)$ & $(b)$
  \end{tabular}
  \caption{A numerical convergence test for a rising gas bubble in a liquid at four different resolutions: $33\times 65$ (black, -), $65\times 129$ (black, - -), $129\times 257$ (red, -), $257\times 513$ (red, - -); $(a)$~the bubble shapes are shown in a domain $(x,z)\in [-1,1]\times [0,4]$ at $t=0,\,2T,\,4T$; $(b)$~the phase field variable $C(0,z,t)$ for $t=2T$ and $4T$. }
  \label{fig:bbct}
\end{figure}

Fig~\ref{fig:bbct}$(a)$ shows a comparison of the shape of the bubble at various resolutions and times. The gas-liquid interface at $t=0$ gives the initial shape of the bubble, which is given by the phase field contour $c(x,y,0)=0.5$.  The time evolution of the interface ({\em e.g.} label curve at $c=0.5$) is demonstrated by comparing the phase field contours between $4$ resolutions (as stated above), and the results are given for $t=2T$ and $t=4T$ in Fig~\ref{fig:bbct}($a$).  Clearly, successively refined meshes show only small differences at bottom and top edges of the bubble with respect to the coarsest resolution $33\times 65$. As expected in the wavelet method, the increase of resolution has increased the accuracy of predicting the gas liquid interface. The rate of convergence for the present method shows an excellent agreement with what was presented by~\citet{Ding2007} (Figs3,~4 therein). Clearly, the resolution $129\times 257$ with $\Delta = 0.0156D$ is sufficient for this test case. The thickness of the interface for this phase field simulation is about $0.07D$, and $4$-$5$ grid points across this interface seems acceptable.

\subsubsection{Effects of entrainment and frictional pressure loss}
The change in bubble's shape at later times is associated with effects of entrainment from the surrounding liquid and the frictional pressure loss. 
When the gas bubble rises due to the buoyancy force, the conservation of mass causes the entrainment of surrounding liquid. As a result, frictional pressure loss is felt in the liquid closed to the bubble, where the vertical pressure gradient is approximately balanced by the viscous stress so that $\frac{\partial p}{\partial z} = \mu\frac{\partial^2w}{\partial z^2}$.  Above the top surface of the gas bubble, entrainment of surrounding liquid leads to $\frac{\partial^2w}{\partial z^2} > 0$. However, in the liquid below the bubble the frictional pressure loss results in $\frac{\partial^2w}{\partial z^2} < 0$. Clearly, a reversed flow exists beneath the bubble due to the adverse pressure gradient of the frictional pressure loss, which  induces a jet that pushes into the bubble from below. Thus, the jet force interacts with the capillary force in the interface where the liquid and the gas meet face-to-face. 

To show how the jet affects the rate of convergence, we present plots of the phase field $c(0,z,2T)$ and $c(0,z,4T)$ in Fig~\ref{fig:bbct}$(b)$, where the oscillations indicate that the corresponding resolution is insufficient for capturing the capillary effect. As the resolution increases, the gas-liquid interface is accurately resolved, and the oscillations have disappeared. At the lower resolutions $33\times 65$ and $65\times 129$, there are noticeable discrepancies between two solutions at both times. However, at higher resolutions $129\times 257$ and $257\times 513$, we have a converged solution. One of the most important aspects of the present simulation is that the interface does not smear at low resolution, which indicates that artificial numerical diffusion is absent in these simulations. 
\subsection{A comparison with experimental investigations}
In this section, we report two experimental results to validate the phase field approach for modelling a two-phase flow. In the first experiment, an air bubble was injected in a water filled vertical cylindrical tube that has a diameter of $11.6$~cm and a length of $100$~cm. A details of the experiment is given by~\citet{Rahman2015}. Initially, a spherical air bubble of volume $7\times 10^{-7}\hbox{ m}^3$ was injected. Average terminal velocity of the air bubble in water was calculated $23.2$~cm/s. 

We have performed a numerical simulation of the above experiment, where the diameter and the length of the tube are $11.6$~cm and $46.4$~cm, respectively. Values of the physical parameters from the experiment have been used to determine all dimensionless constants of the numerical simulation. We have also performed a numerical convergence test using $6$ meshes of different resolutions: $17\times 65$, $33\times 129$, $65\times 129$, $65\times 257$, and $129\times 513$. Based on the converged solution, the terminal velocity of the rising air bubble is found $23.9$~cm/s. The discrepancy between the experimental and the numerical results is about $2.5$\%. As stated by~\citet{Rahman2015}, the measurement accuracy must also be considered when the experimental results are compared with the numerical results. Thus, we consider that the above comparison a very good indication to be convinced that the present Allen-Cahn phase field method models a gas-liquid two-phase flow accurately at least when a single bubble rises in a quiescent liquid.

 \begin{figure}
   \centering
   \includegraphics[width=6cm]{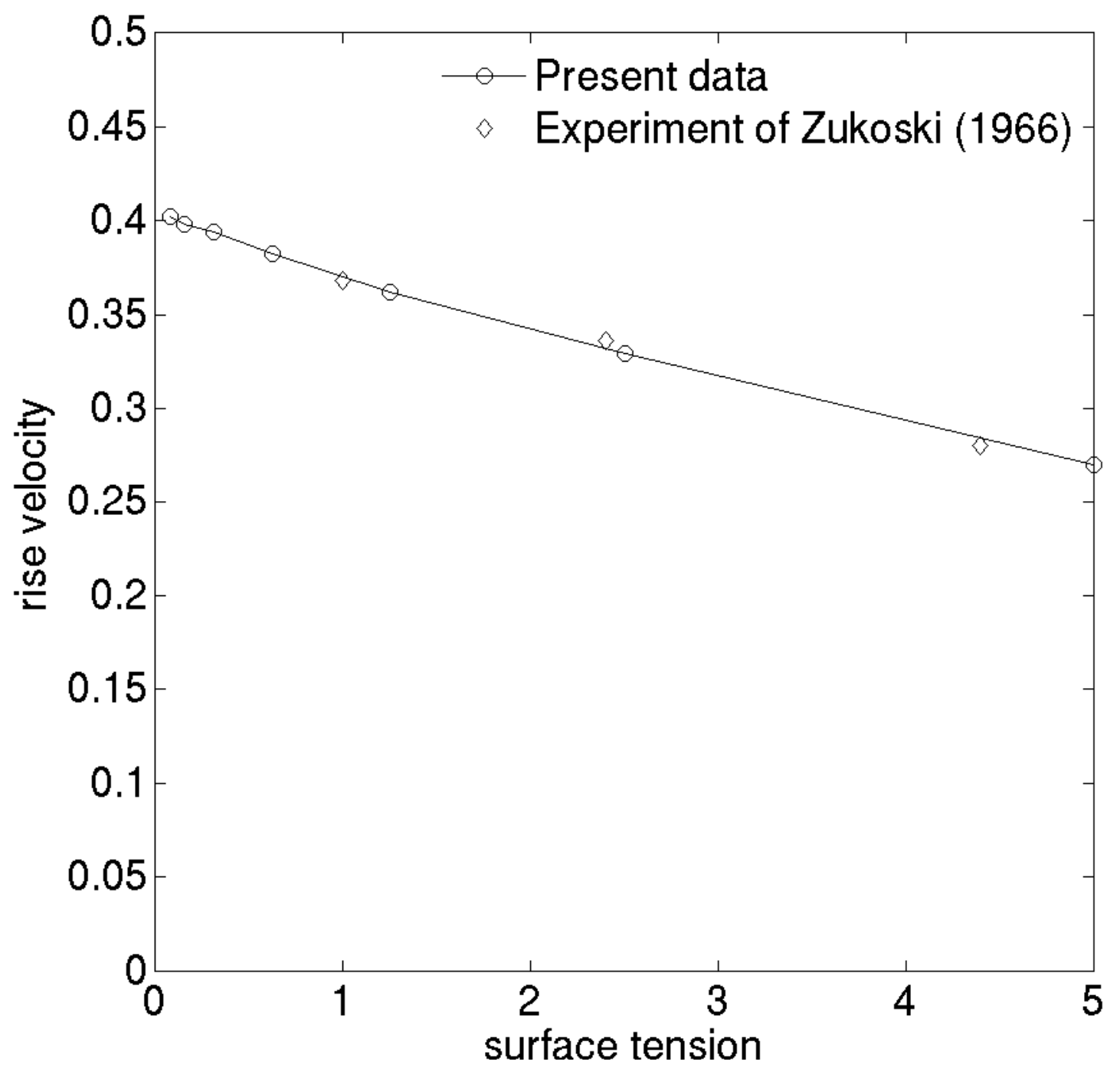}  
   \caption{A comparison for the relationship between the surface tension parameter $\sigma/\rho_l W^2 D$ and the rise velocity $w/W$. Data from the present simulation is compared with that from the experiment of~\citet{Zukoski66}.}
   \label{fig:sgw}
 \end{figure}

A details of the second experiment was given by~\citet{Zukoski66}, where the influence of surface tension on the bubble's terminal velocity was investigated (e.g. see Table 1 therein). We have normalized the surface tension force $\sigma/D$ per unit area by $\rho_lU^2$. The dimensionless bubble rising velocity $w/W$ is then computed for various values of the normalized surface tension parameter $\sigma/\rho_l W^2 D$. Our numerical simulation indicates that an increase of surface tension decreases the rise velocity, and the result is consistent with the experimental data reported by~\citet{Zukoski66}. For a comparison, the variation of rise velocity ($w/W$) with respect to the surface tension parameter ($\sigma/\rho_l W^2 D$) is presented in Fig~\ref{fig:sgw}. The result has to be interpreted carefully. The agreement between the present simulation and the experiment -- as depicted in Fig~\ref{fig:sgw} -- shows the accuracy of the present method of implementing surface tension force on the right hand side of~(\ref{eq:nse}). Note that surface tension force has been calculated with the method demonstrated by~\citet{Jacqmin99} and ~\citet{Ding2007} based on the Cahn-Hilliard phase field method. We have calculated the surface tension force using the Allen-Chan phase method. 

\subsection{The instability of a rising gas bubble}
Let us briefly summarize our investigation on the stability of a rising bubble.
As seen in Fig~\ref{fig:bbct}, the interface -- separating the liquid above from the gas below -- of such a bubble is smooth and closely spherical, although ripples may be seen under some conditions~\cite{Davies50,Batchelor87,Hua2007}. For example, if such an interface were planner, it might exhibit instability at disturbances of wavelength exceeding a critical value $\lambda_c=2\pi\sqrt{\frac{\sigma}{\Delta\rho g}}$, which is approximately $1.7$~cm for the air-water interface. Among several factors for stability, the time scale for the growth of the disturbance may be swept by the entrainment of liquid flow. In Fig~\ref{fig:bbct}$a$, one observes that spherical (2D) shape of the bubble is deformed to a skirted bubble at $t=4T$; however, the upper surface of the bubble remains spherical. In~\cite{Hua2007} the effect of surface tension and the viscosity was investigated and no instability on the uppers surface of a rising bubble was observed for at least $\mathcal Re\le 200$. %The upper surface of a larger spherical bubble may be gravitationally unstable, and such bubbles may break up into smaller bubbles with stable interfaces~\cite{Davies50,Batchelor87,Bhaga81}. 

\begin{figure}
  \centering
  \begin{tabular}{cc}
    \multicolumn{2}{c}{
    \includegraphics[height=4.5cm]{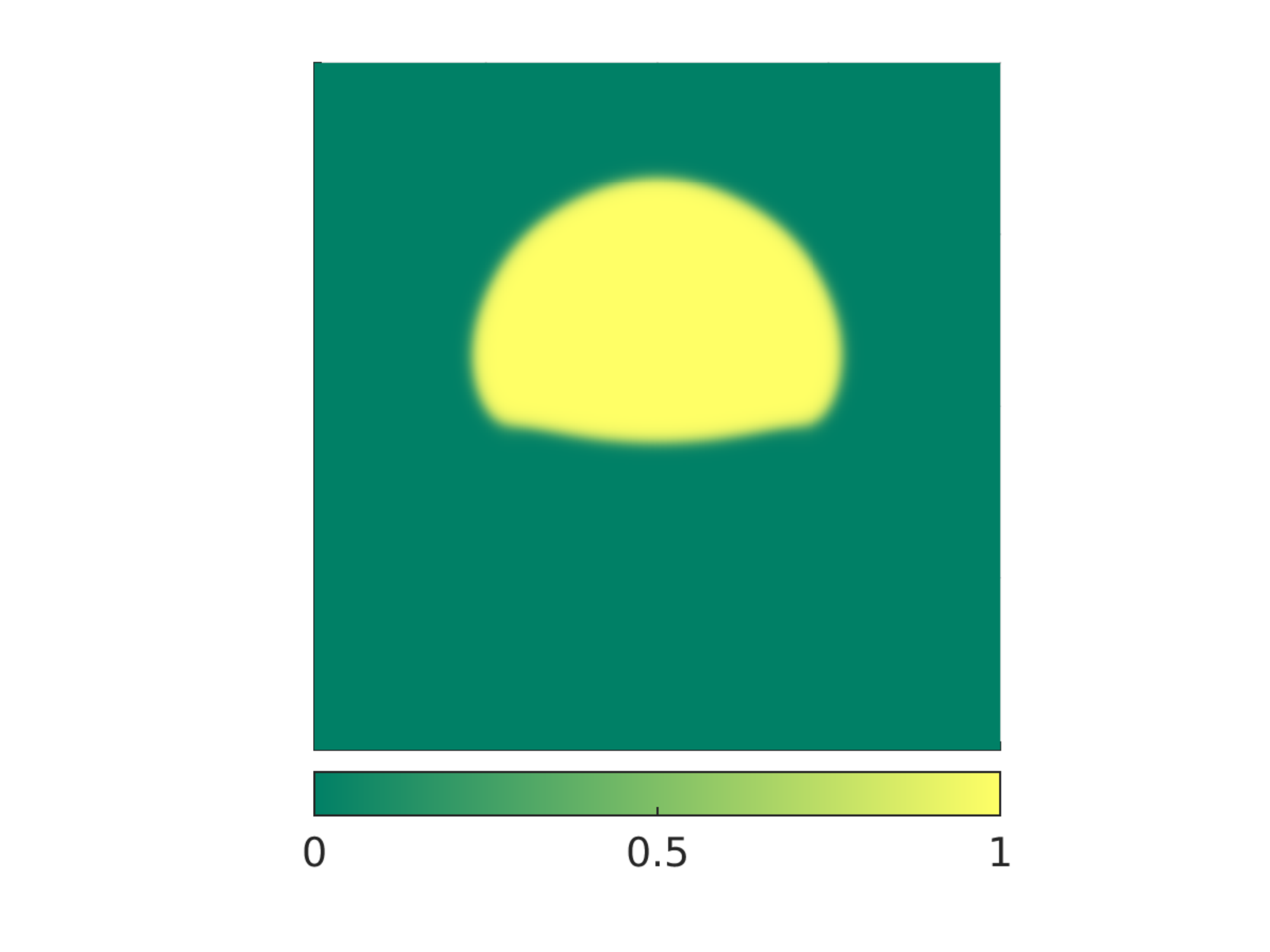}
    }\\
    \multicolumn{2}{c}{
      %$(a)$ Cahn = $0.0316$
      $(a)$ $\mathcal Pe = 10^3$
    }\\
      \includegraphics[height=3.5cm]{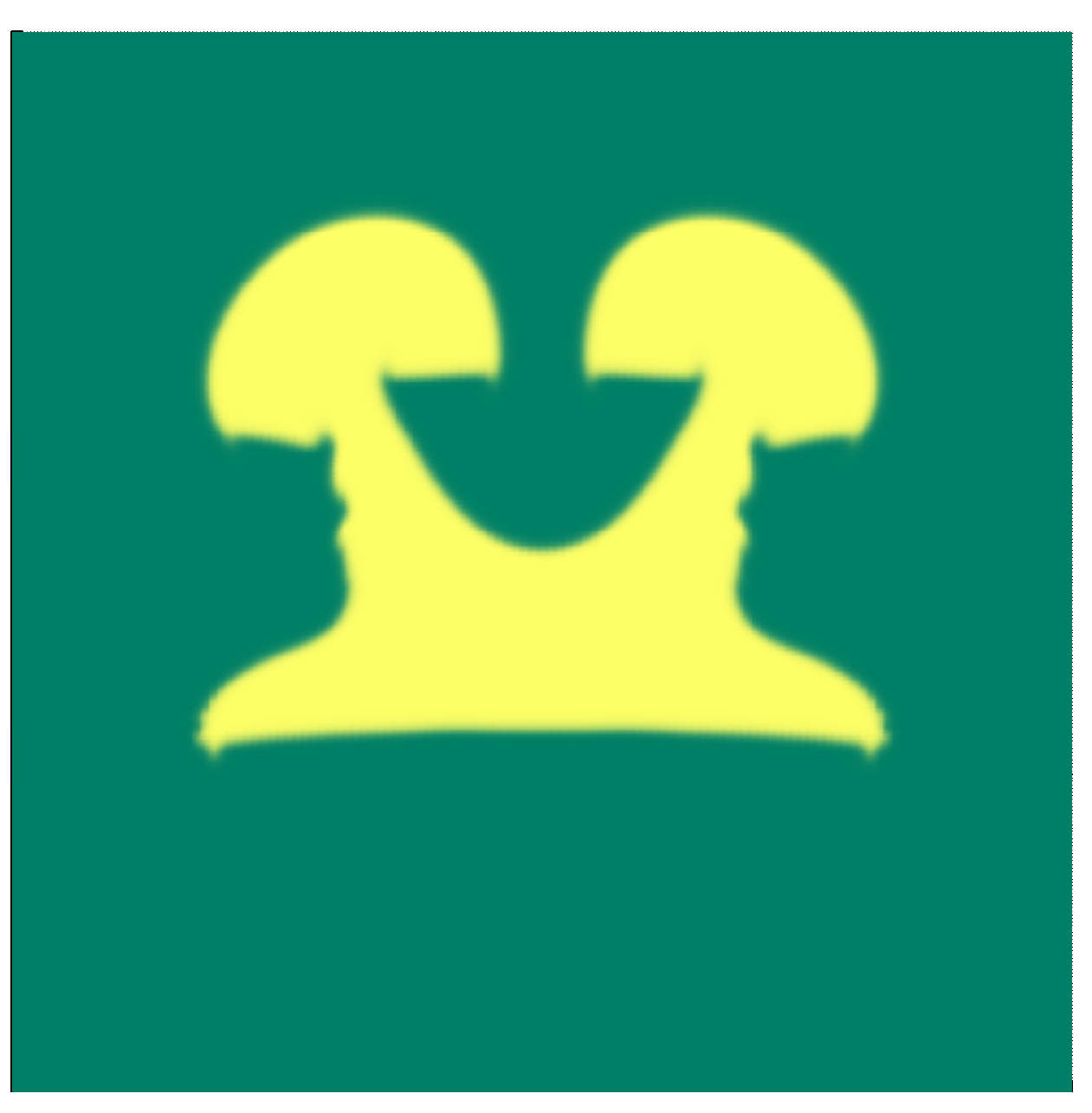}
    &
      \includegraphics[height=3.5cm]{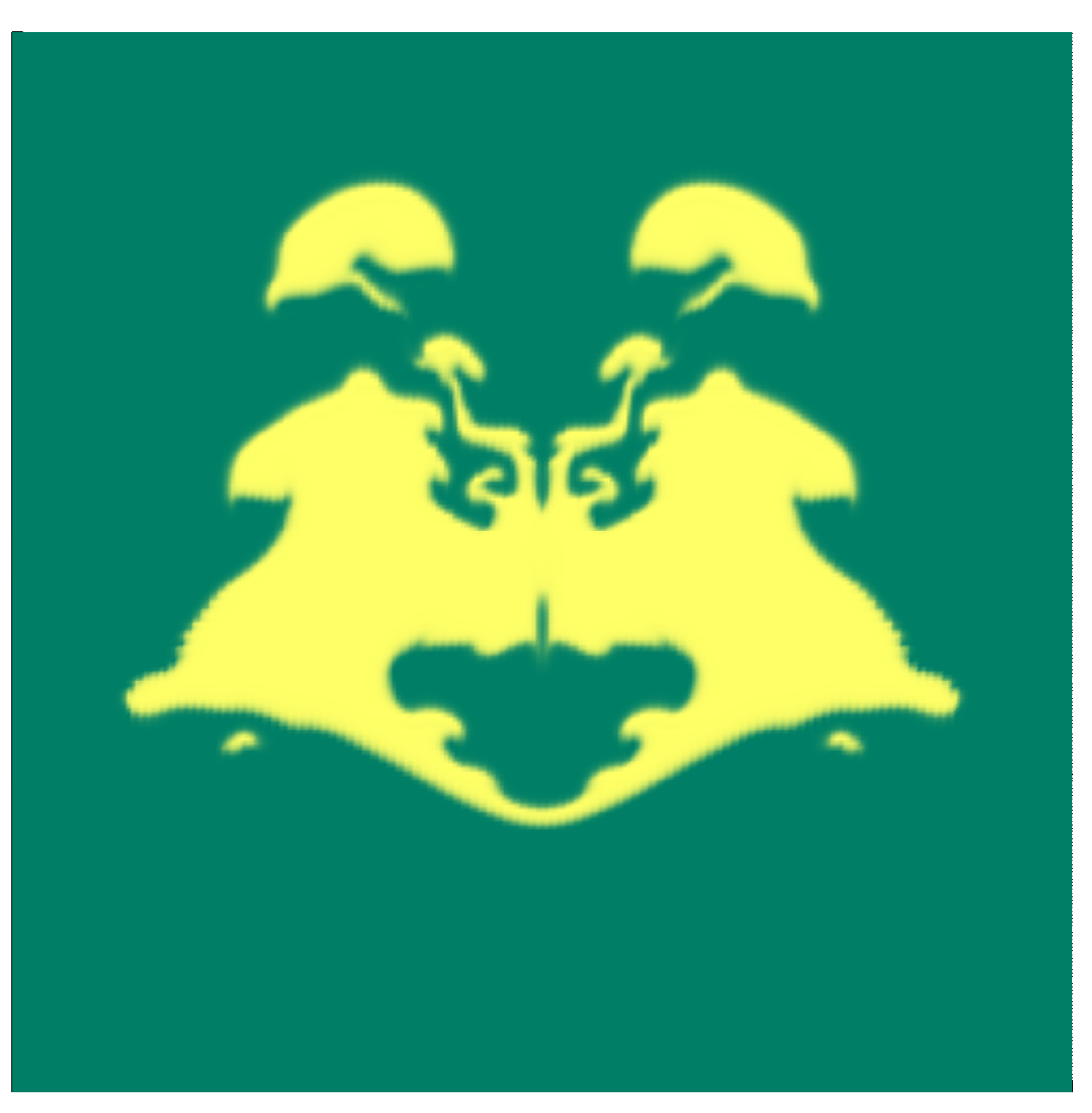}
    \\
    $(b)$ $\mathcal Pe = 4\times 10^3$ & $(c)$ $\mathcal Pe=8\times 10^3$ \\
    %$(b)$ Cahn = $0.0158$ & $(c)$ Cahn = $0.0112$ \\
    \includegraphics[height=3.5cm]{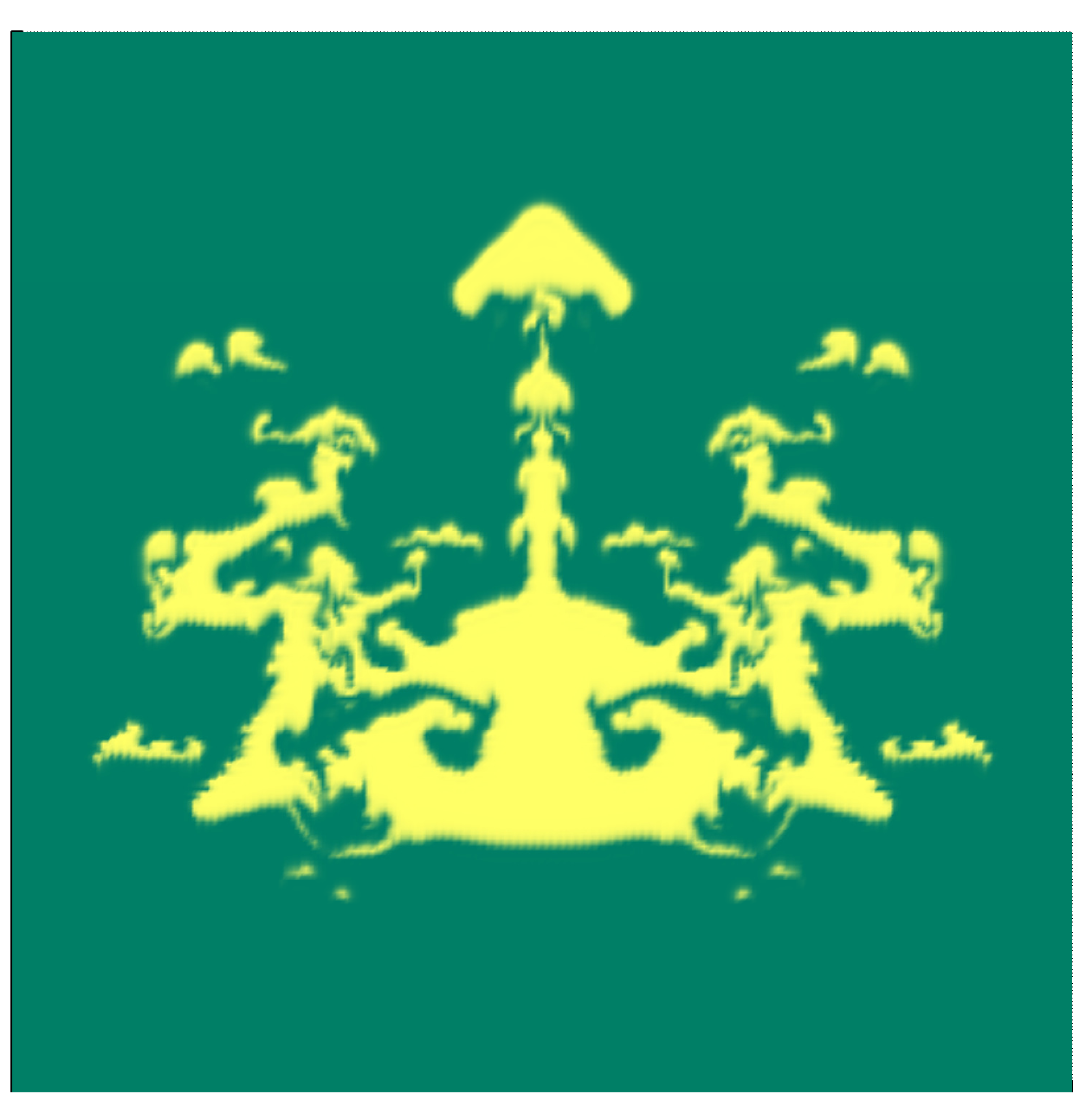}
    &
      \includegraphics[height=3.5cm,width=3.5cm]{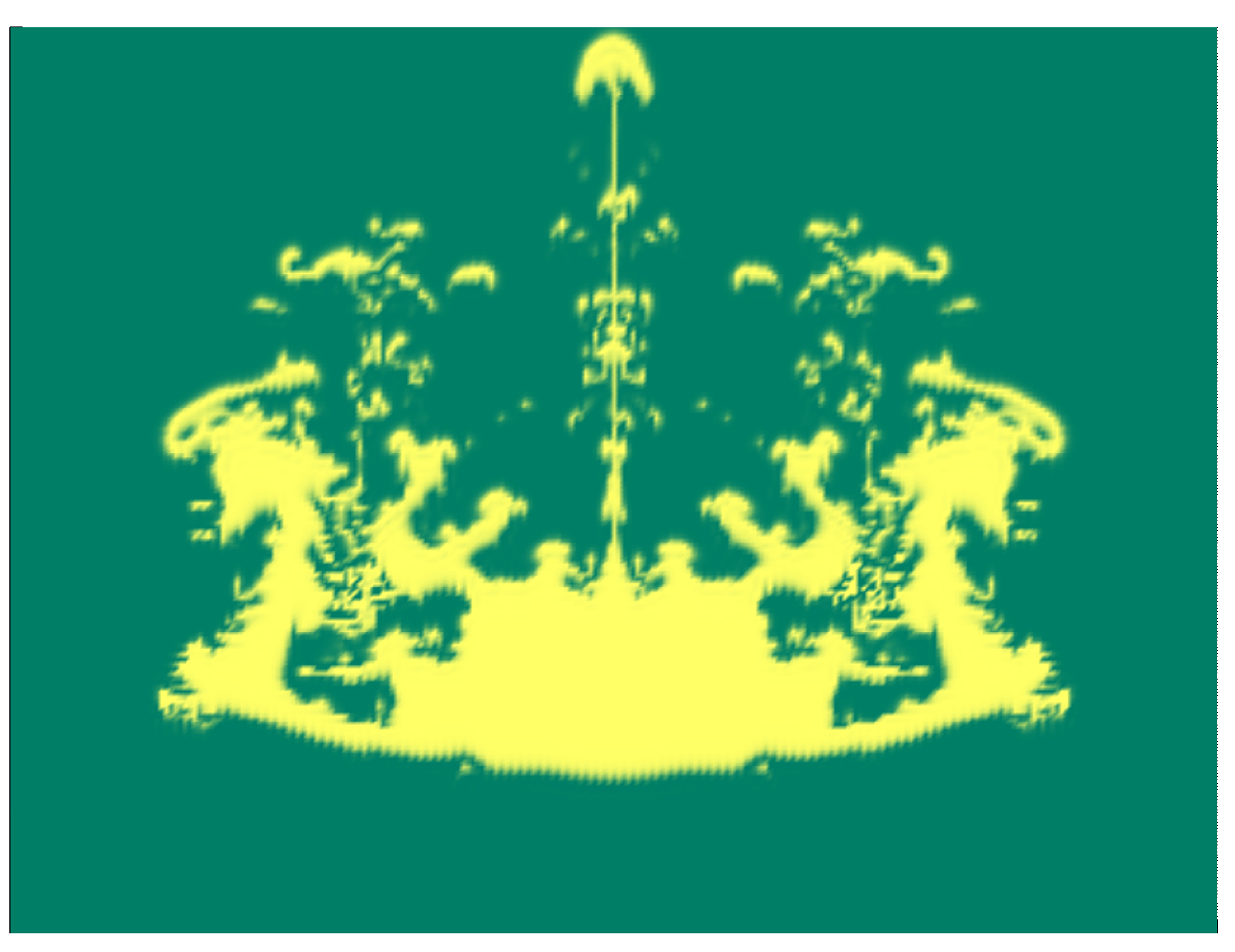}
    \\
    $(d)$ $\mathcal Pe = 1.6\times 10^4$ & $(e)$ $\mathcal Pe=3.2\times 10^4$ \\
    %$(d)$ Cahn = $0.0079$ & $(e)$ Cahn = $0.0056$ 
  \end{tabular}
  \caption{The instability of a rising bubble with respect to increasing $\mathcal Pe$. The initial condition is the same as what is in case of Fig~\ref{fig:bbct}.}
  \label{fig:fngr}
\end{figure}

In the phase field simulation, the Cahn number ($\epsilon/D$) characterizes the competition between the hydrophilic and the hydrophobic interaction. For the Allen-Cahn phase field method, the diffusive time scale and convective time scale are given by $\mathcal M\epsilon^2/D^2$ and $W/D$, respectively. The ratio of the diffusive time scale to the convective time scale is the Peclet number $\mathcal Pe$. For a given Cahn number, the larger the Peclet number the slower the diffusion. When there is an imbalance between the two time scales, ripples may occur on the fluid-fluid interface or bubbles may break up into smaller bubbles~\cite{Davies50,Batchelor87,Bhaga81}. Here, we consider $5$ values of the Peclet number -- $10^3$, $4\times 10^3$, $8\times 10^3$, $1.6\times 10^4$, and $3.2\times 10^4$ -- in order to demonstrate the instability of a rising gas bubble.
%$0.0316$, $0.0158$, $0.0112$, $0.0079$, and $0.0056$.
The parameters for this simulation are the same as that listed in table~\ref{tab:pp} except  $D=2.5$~m. The choice for a relatively large value of $D$ aims at achieving a strong adverse pressure gradient due to the frictional pressure loss so that the wavelength of the disturbance exceeds the critical value $\lambda_c$. Plots of the color filled phase field contours at the same reference time corresponding to each Peclet number are shown in Fig~\ref{fig:fngr}. When the Peclet number is $4\times 10^3$, the liquid right above the bubble penetrates directly into the gas taking a shape of mushroom. The entrained liquid around the bubble also penetrates  horizontally into the bubble. The mushroom line penetration of liquid and gas into each other has enhanced at $\mathcal Pe=8\times 10^3$. A further increase of the Peclet number exhibits an unstable but coherent deformation of the initial bubble into smaller mushroom like structures.

\subsection{Rayleigh-Taylor instability}

The Rayleigh-Taylor instability~(RTI) is a well documented buoyancy driven fluid flow that occurs when a lower density fluid is accelerated into a higher density fluid. According to the linear stability theory, perturbations of different scales grow in time independently of one another. In general, disturbances grow like $h = \alpha A h_b$, where $h_b=gt^2$ is the characteristic length scale associated with frictionless up-welling or down-welling, $g$ is acceleration due to gravity, $\mathcal A = (\rho_l-\rho_g)/(\rho_l+\rho_g)$ is the Atwood number, and $\alpha$ is a growth constant. Experiments and numerical simulations show that $\alpha\sim 0.04-0.08$. We now discuss the numerical simulation of RTI using the proposed wavelet based phase field method.

\subsubsection{A numerical verification}
\begin{figure}
  \centering
  \includegraphics[width=8cm,height=8cm]{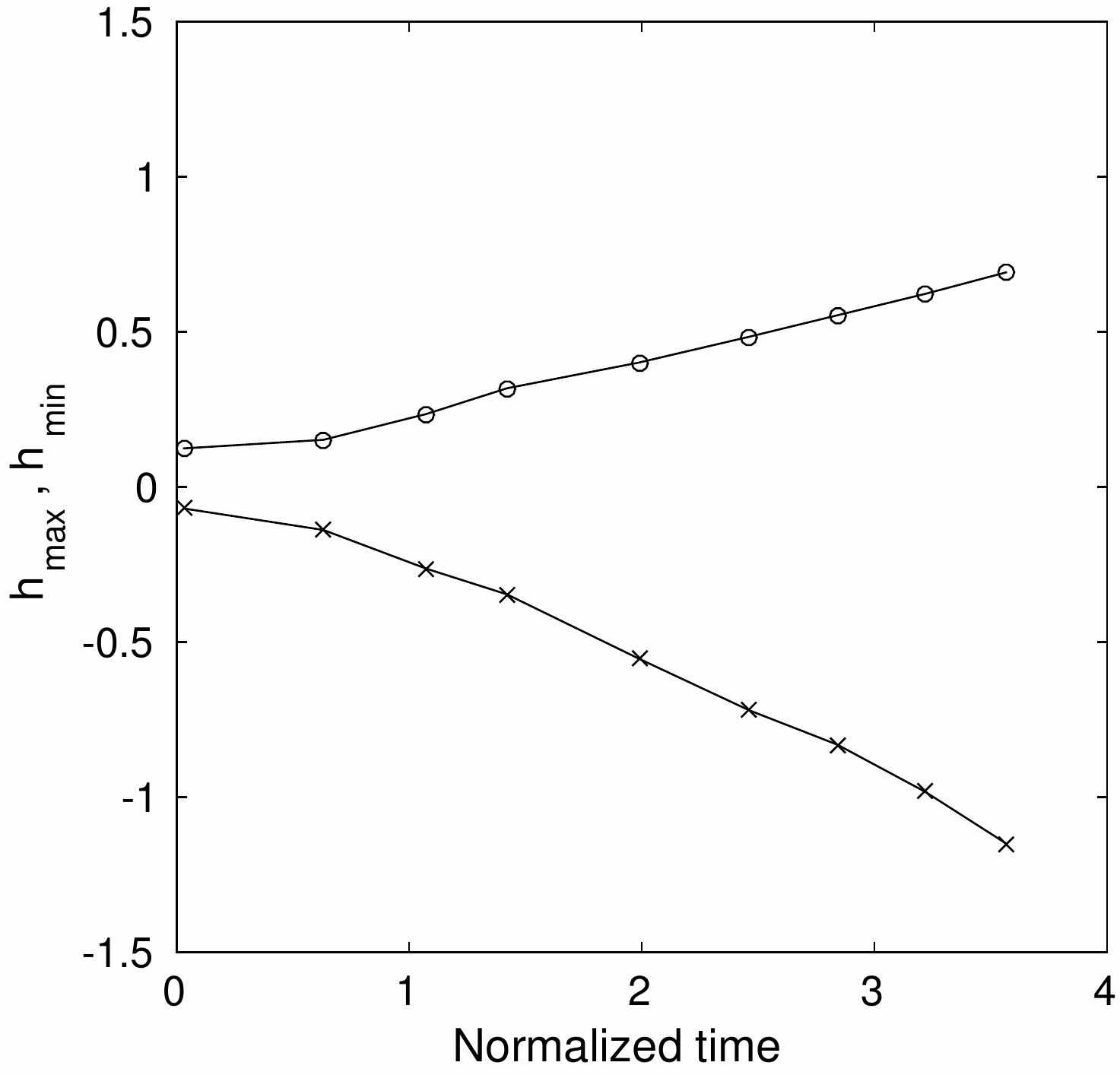}
  \caption{The $z$-coordinate for up-welling ($h_{\max}$, {\tiny o}) and down-welling ($h_{\min}$, {\tiny x}) of the interface defined by $c(x,z,t)=0.5$. }
  \label{fig:rtih}
\end{figure}

A single mode growth of the RTI has been simulated in a rectangular domain $[0,1]\times [0,4]$ at a fixed Reynolds number $Re=\rho_lWD/\mu_l$ for which a number of reference results are available in the literature. The characteristic velocity is given by $W=Fr\sqrt{AgR/(1+A)}$ with $0<Fr<1$. The values of the physical parameters given in table~\ref{tab:pp} except now $\rho_g=333\hbox{ kg/m}^3$. To help comparison with the RTI simulation of~\citet{Ding2007}, we have used $Re\sim 3\,001$ and $A\sim 0.5$, where $\mu_g=\mu_l$ and $\rho_g/\rho_l\sim 0.333$. 
As expected, the lighter fluid penetrates into the region of heavy fluid and the heavy fluid falls into the lighter fluid. The initial interface is defined by the phase field $0.5-\epsilon/2\le c(x,z,t) \le 0.5+\epsilon/2$, which elongates in time and rolls up into counter rotating vorticies. Up-welling ($h_{\max}$) and down-welling ($h_{\min}$) disturbances of the interface have been obtained such that $h_{\max}(t)$ is the maximum value of $z$, and $h_{\min}(t)$ is the minimum value of $z$ such that $c(x,z,t)=0.5$ at any time $t$. When the plot in Fig~\ref{fig:rtih} is compared with that in Fig~$5$ of~\citet{Ding2007}, one observes an excellent agreement between two results. Readers are also warned that some discrepancies between two data sets are noticeable, which are associated with the difference in the governing equations and the numerical method used between two simulations. It is worth mentioning that for the RTI simulation, as the interface between gas and liquid evolves in time, the vortex sheet roll-up is dependent on the physical properties of two fluids. and the balance between the inertia and viscous effect. For a given Atwood number, an investigation on the roll-up of vortex sheet for a reduced viscosity ratio, $\mu_g/\mu_l$, may help to understand the present simulation method.  

\subsubsection{Viscous effect on RTI at  a fixed  Atwood number, $A=0.5$}
\begin{figure}
  \centering
  \begin{tabular}{cccccc}
    \includegraphics[width=2cm]{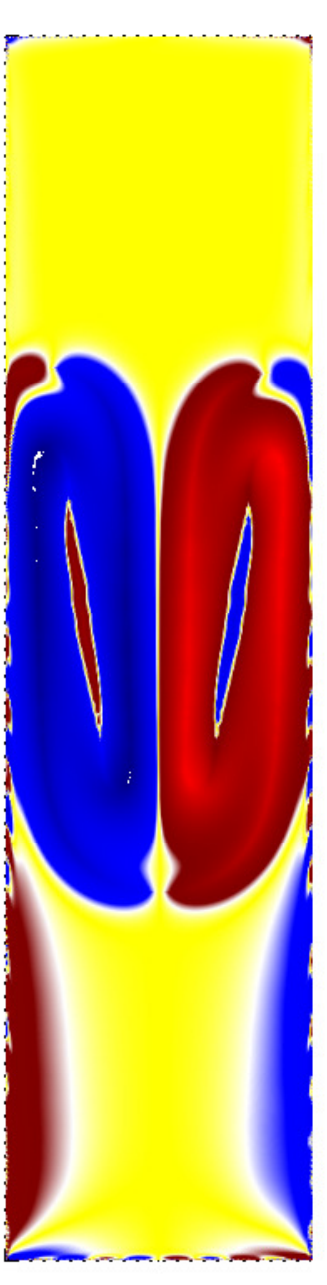}
    &
    \includegraphics[width=2cm]{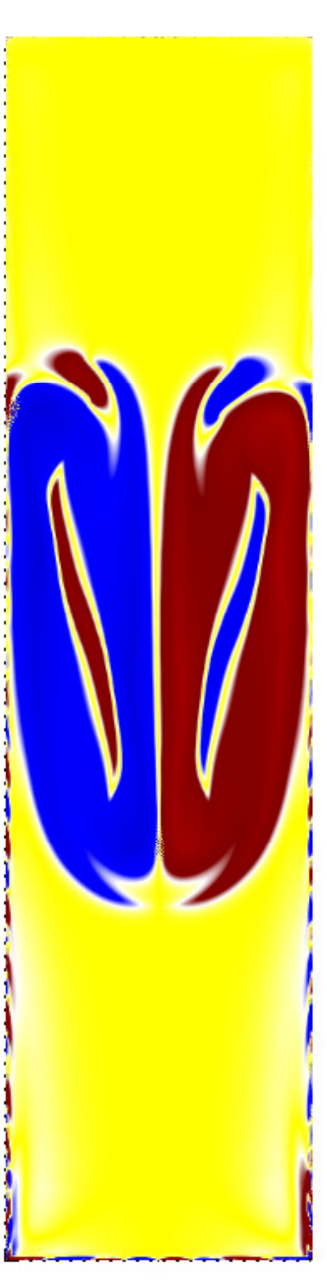}
    &
    \includegraphics[width=2cm]{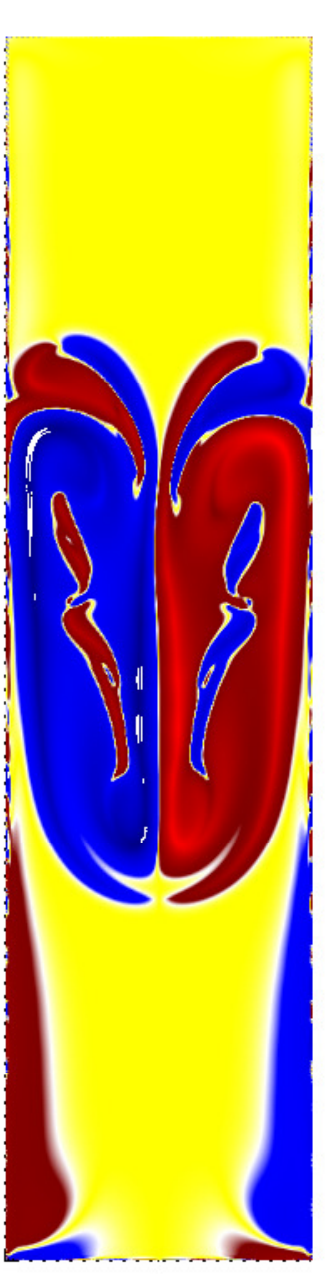}
    &
    \includegraphics[width=2cm]{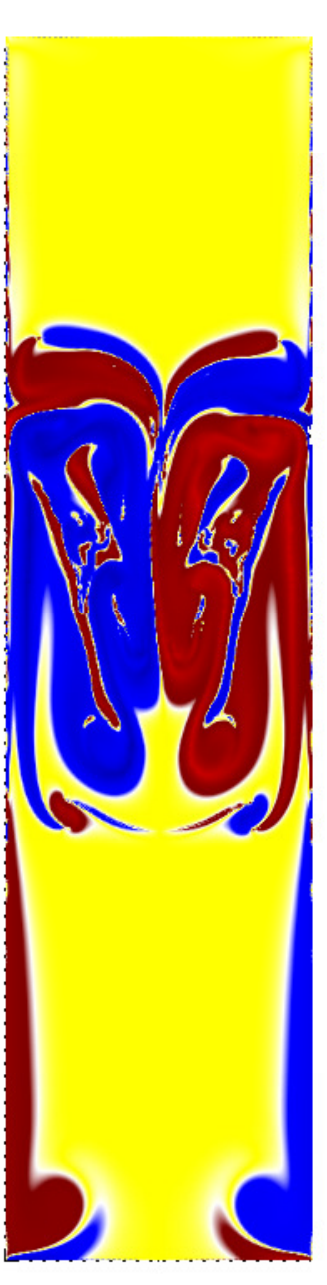}
    &
    \includegraphics[width=2cm]{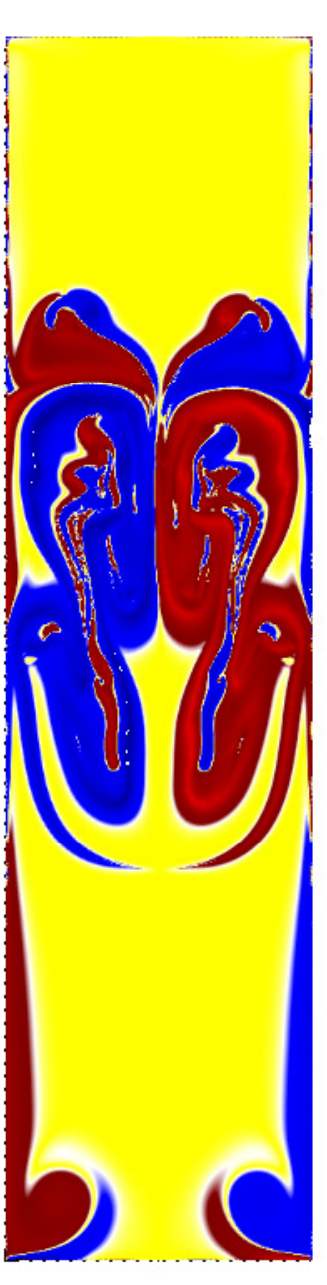}
    &
    \includegraphics[width=2cm]{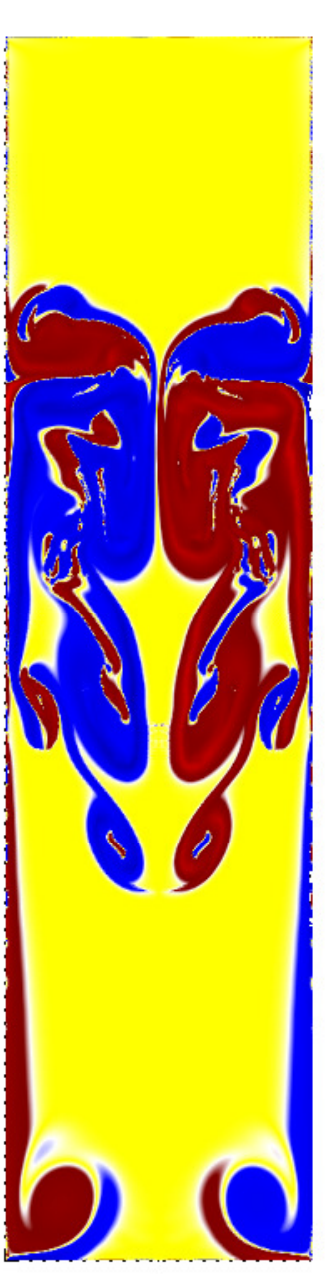}\\
    \includegraphics[width=2cm]{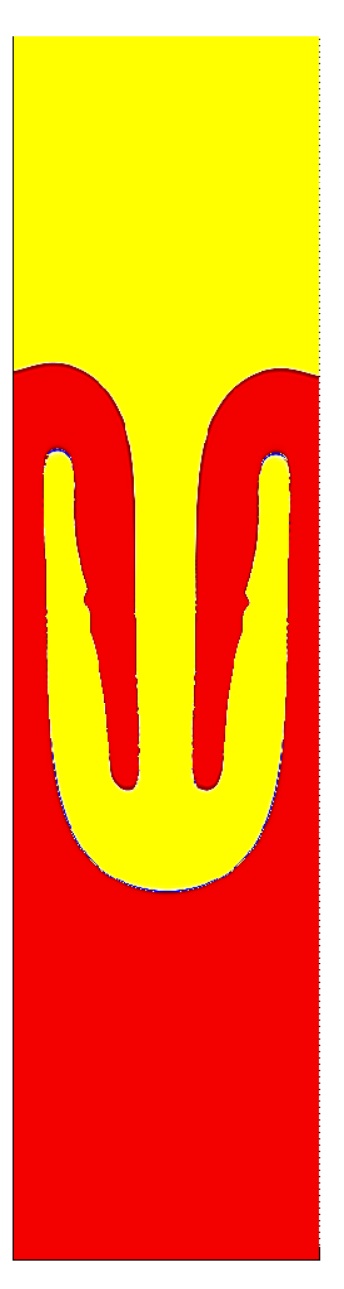}
    &
    \includegraphics[width=2cm]{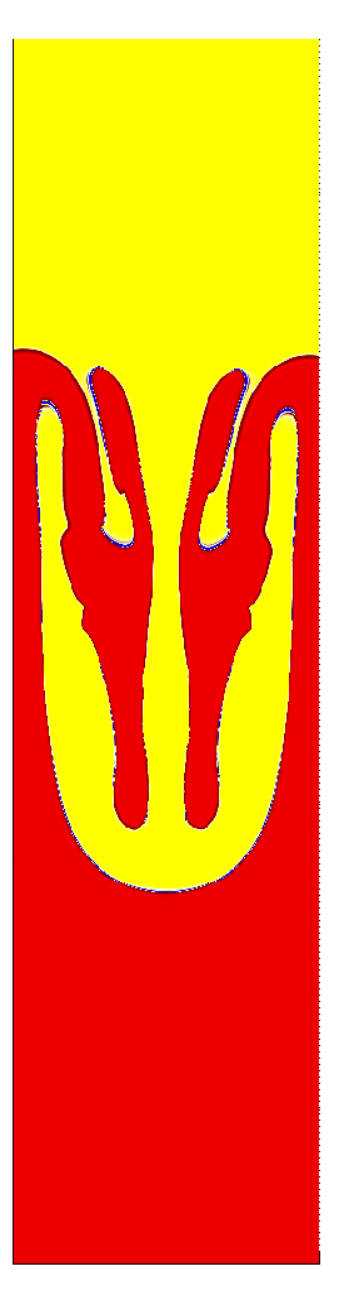}
    &
    \includegraphics[width=2cm]{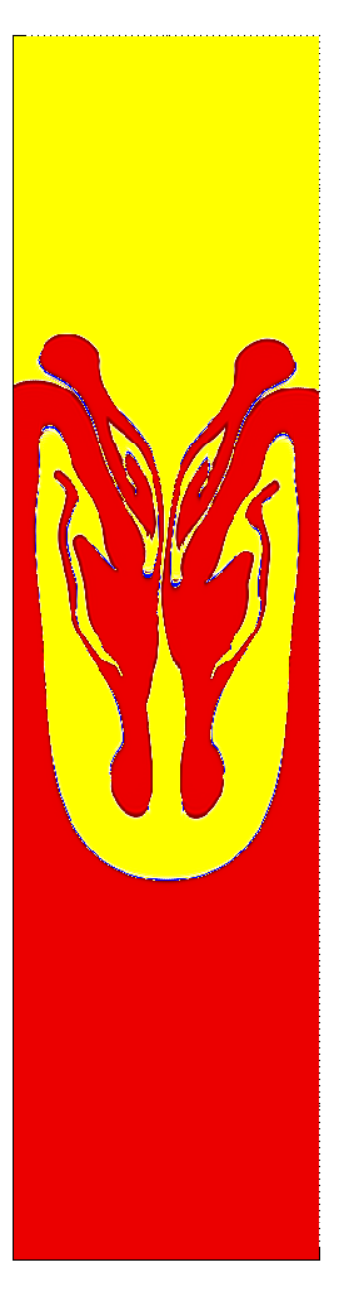}
    &
    \includegraphics[width=2cm]{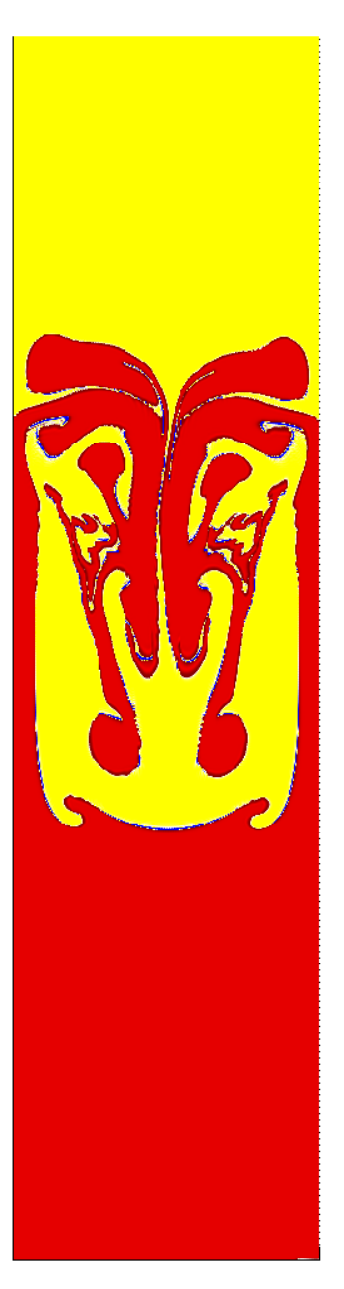}
    &
    \includegraphics[width=2cm]{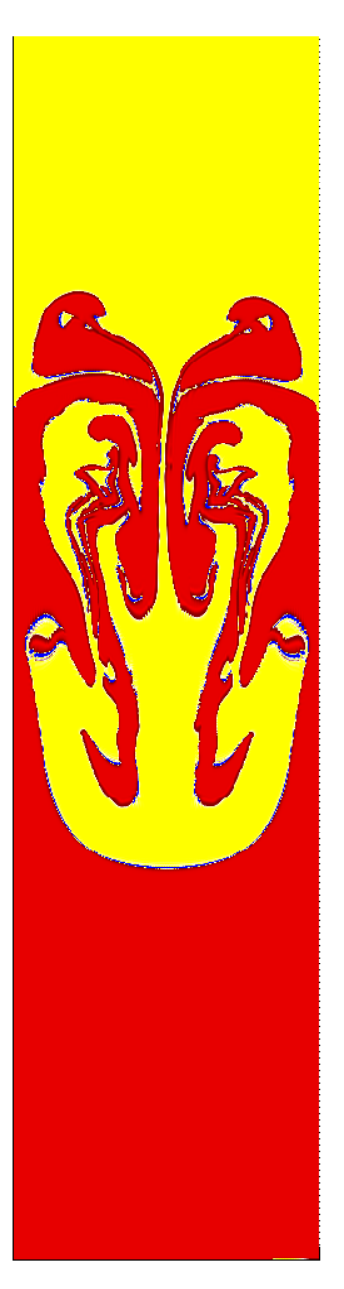}
    &
    \includegraphics[width=2cm]{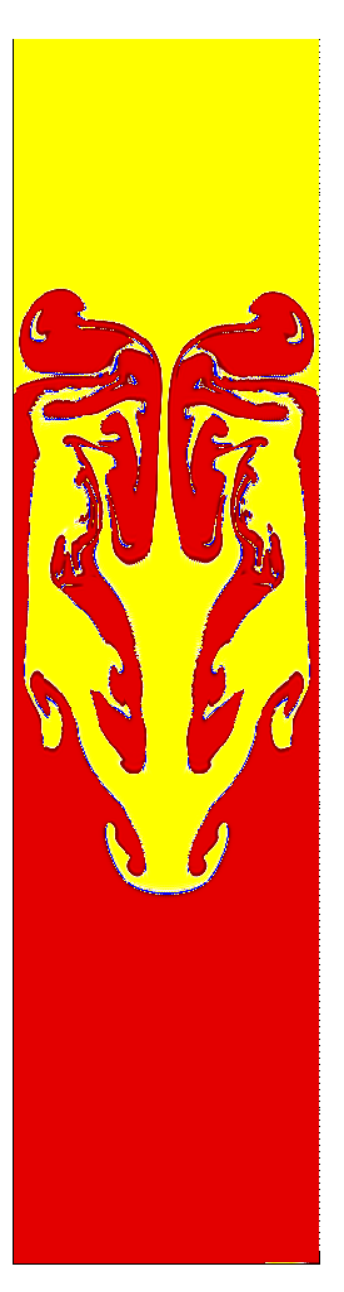}\\
    %$Re=1\,000$ & $Re=2\,000$ & $Re=4\,000$ & $Re=8\,000$ & $Re=10\,000$ & $Re=12\,000$\\
    %$\mu_g/\mu_l=1$ & $\mu_g/\mu_l=0.5$ & $\mu_g/\mu_l=0.25$ & $\mu_g/\mu_l=0.125$ & $\mu_g/\mu_l=0.1$ & $\mu_g/\mu_l=0.083$\\
    $\mu_l/\mu_g=1$ & $\mu_l/\mu_g=2$ & $\mu_l/\mu_g=4$ & $\mu_l/\mu_g=8$ & $\mu_l/\mu_g=10$ & $\mu_l/\mu_g=12$
  \end{tabular}
  \caption{A demonstration of the Rayleigh Taylor instability. Top row: regions of counter clockwise (red) and clockwise (blue) vorticity are shown, where yellow indicates a zero vorticity. Creation and roll up of vorticity filaments are presented as $\mu_g/\mu_l$ increases from left to right. Bottom row: shaded plots of the phase field ($c$), where regions of $c\sim 1$ (red) and $c\sim 0$ (yellow) are shown.}
  \label{fig:rtic}
\end{figure}
For small amplitude perturbations of an idealized gas-liquid interface, inviscid linear theory suggests an unbounded exponential growth, where different scales grow independently of one another. However, viscous effect may prevent from high wavenumber growth~\cite{Batchelor87,Tryggvason2001,Jeffrey2012,Zukoski66}; as a result, a `most unstable' mode of solution is produced, which may be seen from the mechanical energy equation below. As the perturbations grow in a two-dimensional RTI, the gas-liquid interface deforms into counter-rotating rolls in the form of vorticity filaments. Note that creation and roll-up of vorticity filaments are characteristics for transition to two-dimensional turbulence. In the absence of viscosity contrast, {\em i.e.} $\mu_g/\mu_l=1$ and $\mu = \mu_gc+(1-c)\mu_l$, perturbations of the interface lead to horizontal buoyancy gradients at the edges of rising gas bubbles, which generate vorticity. Near the left side, the clockwise vorticity is due to the positive horizontal buoyancy gradient. Near the right side, the counter-clockwise vorticity is due to the negative horizontal buoyancy gradient. Clearly, viscosity contrast reduces viscous dissipation for a given strain rate tensor. As a result, phase relationship between the horizontal and the vertical velocities is altered, which leads to a gain in the overall kinetic energy.  In order to test the destabilization due to viscosity contrast between oil and gas, we have simulated RTI for different values of the ratio $\mu_l/\mu_g$. 
Let us summarize results from our numerical simulation, where $\mu_g$ is decreased such that $\mu_l/\mu_g=2$, $\mu_l/\mu_g=4$, $\mu_l/\mu_g=8$, $\mu_l/\mu_g=10$, and $\mu_l/\mu_g=12$. Fig~\ref{fig:rtic} demonstrate the creation and roll-up of vorticity filaments, where the gas-liquid interface is also presented as color filled contour plots of the phase field variable $c(x,z,t)$.  %As seen below, a change in $\mu_l/\mu_g$ introduce a change in the effective viscous dissipation in the interfacial region, and hence, the phase relationship between the horizontal~($u$) and the vertical~($w$) velocities is altered. This implies the shear production of energy to dominate over the viscous dissipation of energy.

Let us examine the effect of viscosity contrast, where
$\mu = \mu_gc + (1-c)\mu_l$. We assume that total velocity $u_i+U_i$ consists of a component, $u_i(\bm x,t)$ that is associated to $\mu_g < \mu_l$, and the other component $U_i(\bm x)$ that is associated to $\mu_g=\mu_l$. The energy equation for this case takes the form
$$\frac{d}{dt}\int\frac{1}{2}u_i^2dV = -\int u_iu_j\frac{\partial U_i}{\partial x_j} dV - \phi,$$
where $\phi$ is the viscous dissipation that is associated with the viscosity contrast. For the disturbance that appears in the present RTI simulation, we may define $U_i=[0,W(z),0]$, where the energy equation takes the following form
$$
\frac{d}{dt}\int\frac{1}{2}(u^2+w^2) d\bm x = -\int uw\frac{\partial W}{\partial z} - \phi.
$$
Note that the stress $uw$ averaged over a period is zero if the velocity components $u$ and $w$ are out of phase of $\pi/2$.   A two phase flow without a point of inflection in the velocity profile and without a viscosity contrast ({\em i.e.} $\phi=0$), the disturbance field cannot extract energy from the basic shear flow unless $u$ and $w$ are out of phase. However, the viscosity contrast changes the phase relationship between $u$ and $w$, which causes the mean value of $-uw\frac{\partial W}{\partial z}$ positive and larger than the dissipation. There is no vortex stretching mechanism in the present two phase RTI, and hence, mean squared vorticity is transferred to smaller scales in the form of vorticity filaments, as depicted in Fig~\ref{fig:rtic}.

\section{Concluding remarks}\label{sec:rmrk}
This article presents a phase field method for the CFD simulation of two-phase incompressible flow. An energetic variational principle is employed to compute the surface tension force. The interfacial dynamics is modelled through the Allen-Cahn-Navier-Stokes equations. The governing equations has been solved with a weighted residual collocation method based on the Deslauriers-Dubuc interpolating wavelets. A rising gas bubble in a liquid is simulated to demonstrate the effective calculation of the surface tension force through the phase field variable, where the results are validated with that from physical experiments. To better understand the proposed method, the instability of a rising gas bubble has been discussed. A numerical simulation is presented showing that the competition between the hydrophilic and hydrophobic interaction leads to many mushroom like structures on the upper surface of a spherical bubble. Since we do not have an opportunity for a one-to-one comparison  with an experiment for the occurrence of such structures in a rising gas bubble, we have simulated the Rayleigh-Taylor instability because this representative test case was investigated by a number of other researchers. The plot shown in Fig~5 has an excellent agreement with what was reported by previous authors. In particular, one may compare this plot with Fig~5 of~\citet{Ding2007}. We have also discussed the effects of viscosity contrast on RTI. Clearly, the present research provides useful feedback on effective numerical simulation of two-phase flow problems.

The present method can be extended to simulate three-dimensional two phase flows. However, in 3D, there is a large number of computational degrees of freedom. We need to understand parallel computation of multiresolution wavelet method. It is important to understand techniques for preconditioning the linearized system that appears in the present Newton-Krylov-Wavelet method. The results presented in the paper suggests that the wavelet based Allen-Cahn phase field method is an effective technique for simulating two-phase flow problems. Further investigation in this direction is currently underway. An efficient but simple wavelet method based on the iterative subdivision scheme has been considered in this article. We expect to investigate some advanced features of the wavelet method in the context of phase field theory, which is something that is important but not fully understood. 

\section*{Acknowledgements}
The author acknowledges the discovery grant from the National Science and Research Council~(NSERC), Canada. JMA thanks Nikolas Provatas, McGill University, Canada for introducing the phase field method and for inspiring the possibility that two-phase flows can be simulated effectively using the phase-field method.  This work was benefited by the computing facility of the Shared Hierarchical Academic Research Computing Network ({\tt SHARCNET:www.sharcnet.ca}) and Compute/Calcul Canada.

\bibliographystyle{model1-num-names}
\bibliography{bibrefs}
\end{document}